%% file: main.tex
\documentclass[sigconf]{acmart}

\usepackage{subcaption}  
\usepackage{booktabs}    
\usepackage{colortbl}    
\usepackage{multirow}    
\usepackage{xcolor}      
\usepackage{tcolorbox}   
\tcbuselibrary{breakable, skins} 
\usepackage{lipsum}      
\usepackage{algorithm}   
\usepackage{algorithmic} 
\usepackage{graphicx}   
\usepackage{amsmath}    
\usepackage{enumitem}   
\usepackage{bbding}     

\definecolor{redtext}{RGB}{217,0,0}
\definecolor{lightpink}{RGB}{255,228,225}  
\definecolor{darkpink}{RGB}{255,182,193}   
\definecolor{checkgreen}{RGB}{34,139,34}   
\definecolor{darkgreen}{RGB}{0,128,0}      
\definecolor{darkyellow}{RGB}{218,165,32}   

\AtBeginDocument{%
  }

\setcopyright{none}
\renewcommand\footnotetextcopyrightpermission[1]{}

\begin{document}

\title{
Modeling Hypergraph Using Large Language Models
 }


\author{Bingqiao Gu}
\affiliation{%
  \institution{School of Airspace Science and Engineering, Shandong University}
  \city{Weihai}
  \postcode{264209}
  \country{China}
}
\email{gubingqiao@mail.sdu.edu.cn}

\author{Xingqin Qi}
\affiliation{%
  \institution{School of Mathematics and Statistics, Shandong University}
  \city{Weihai}
  \postcode{264209}
  \country{China}
}
\email{qixingqin@sdu.edu.cn}

\author{Jiale Zeng}
\affiliation{%
  \institution{School of Airspace Science and Engineering, Shandong University}
  \city{Weihai}
  \postcode{264209}
  \country{China}
}
\email{zengjiale@mail.sdu.edu.cn}

\author{Dong Li}
\authornote{Corresponding author.}
\affiliation{%
  \institution{School of Airspace Science and Engineering, Shandong University}
  \city{Weihai}
  \postcode{264209}
  \country{China}
}
\email{dongli@sdu.edu.cn}

\renewcommand{\shortauthors}{Gu et al.} 

\begin{abstract}

Due to the advantages of hypergraphs in modeling high-order relationships in complex systems, they have been applied to higher-order clustering, hypergraph neural networks and computer vision. These applications rely heavily on access to high-quality, large-scale real-world hypergraph data. Yet, compared to traditional pairwise graphs, real hypergraph datasets remain scarce in both scale and diversity. This shortage significantly limits the development and evaluation of advanced hypergraph learning algorithms.
Therefore, how to quickly generate large-scale hypergraphs that conform to the characteristics of real networks is a crucial task that has not received sufficient attention. Motivated by recent advances in large language models (LLMs)—particularly their capabilities in semantic reasoning, structured generation, and simulating human behavior—we investigate whether LLMs can facilitate hypergraph generation from a fundamentally new perspective.
We introduce HyperLLM, a novel LLM-driven hypergraph generator that simulates the formation and evolution of hyperedges through a multi-agent collaboration. The framework integrates prompts and structural feedback mechanisms to ensure that the generated hypergraphs reflect key real-world patterns. Extensive experiments across diverse datasets demonstrate that HyperLLM achieves superior fidelity to structural and temporal hypergraph patterns, while requiring minimal statistical priors. Our findings suggest that LLM-based frameworks offer a promising new direction for hypergraph modeling.
  
\end{abstract}

%




\maketitle

\input{Body/1.tex}  

\begin{figure*}[!t]
  \centering
	\caption{
		\textbf{HyperLLM generates realistic hypergraphs}. The top row displays seven structural and dynamic patterns from a real-world email hypergraph. The bottom row shows that our model, HyperLLM, successfully reproduces all seven patterns.
	}
  \label{fig:complete}
	\scalebox{0.82}{
		\begin{tabular}{c|ccccccc}
			\toprule
			\multicolumn{1}{c}{} &
			\multicolumn{5}{c}{\textsc{Structural Patterns}} &
			\multicolumn{2}{c}{\textsc{Dynamic Patterns}}	
			\\
			\cmidrule(lr){2-6}
			\cmidrule(lr){7-8}
			\multicolumn{1}{c}{} &
			Degrees &
			Hyperedge Sizes &
			Intersection Sizes &
			Singular Values &
			Group Degree &
      Intersecting Pairs &
			Power-law persistence
			\\
			\cmidrule(lr){2-6}
			\cmidrule(lr){7-8}
			\rotatebox[origin=l]{90}{Real Data}
			&
			\includegraphics[height=0.762in]{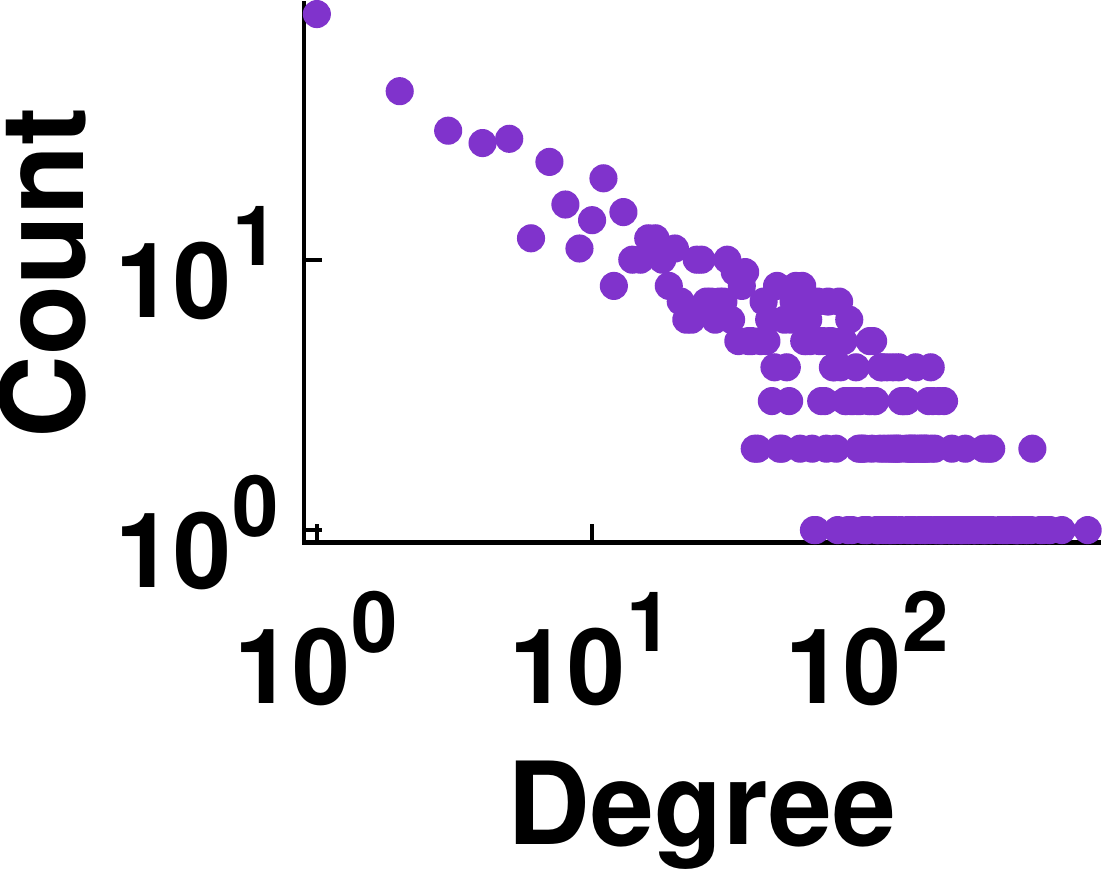} &
			\includegraphics[height=0.762in]{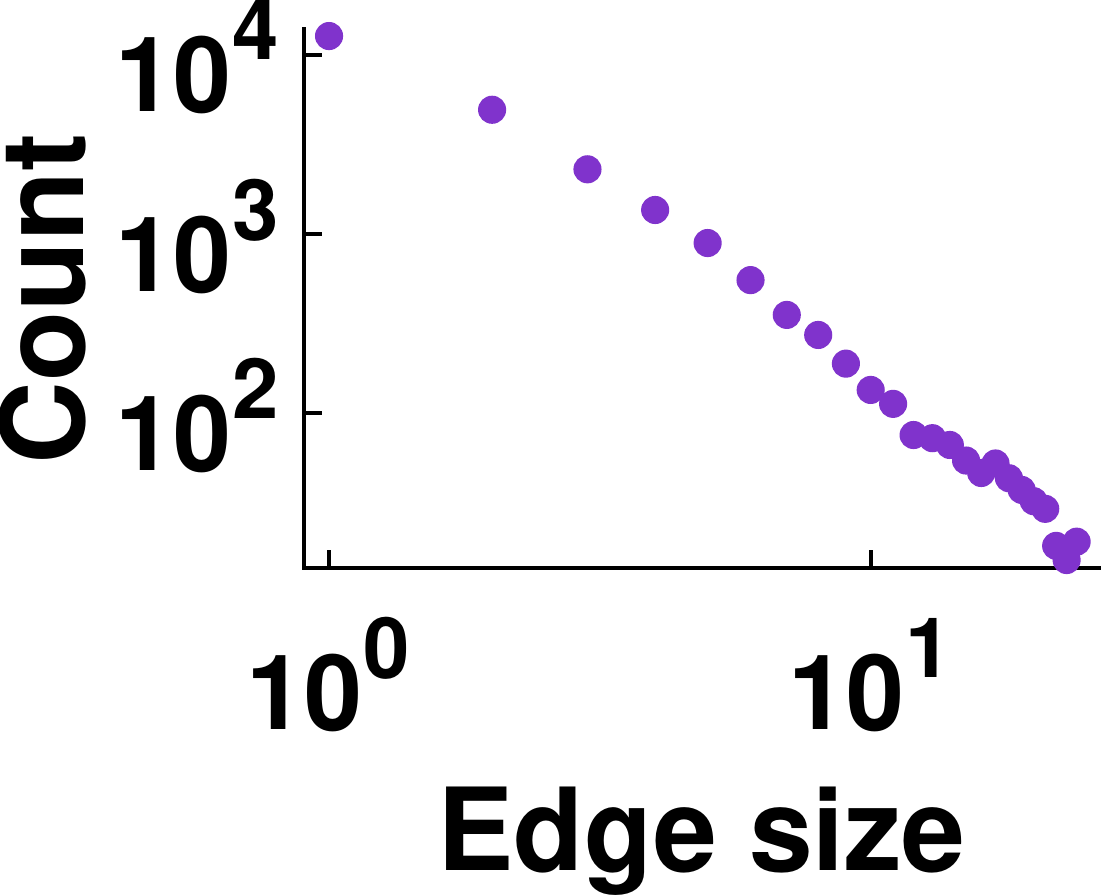} &  	
			\includegraphics[height=0.762in]{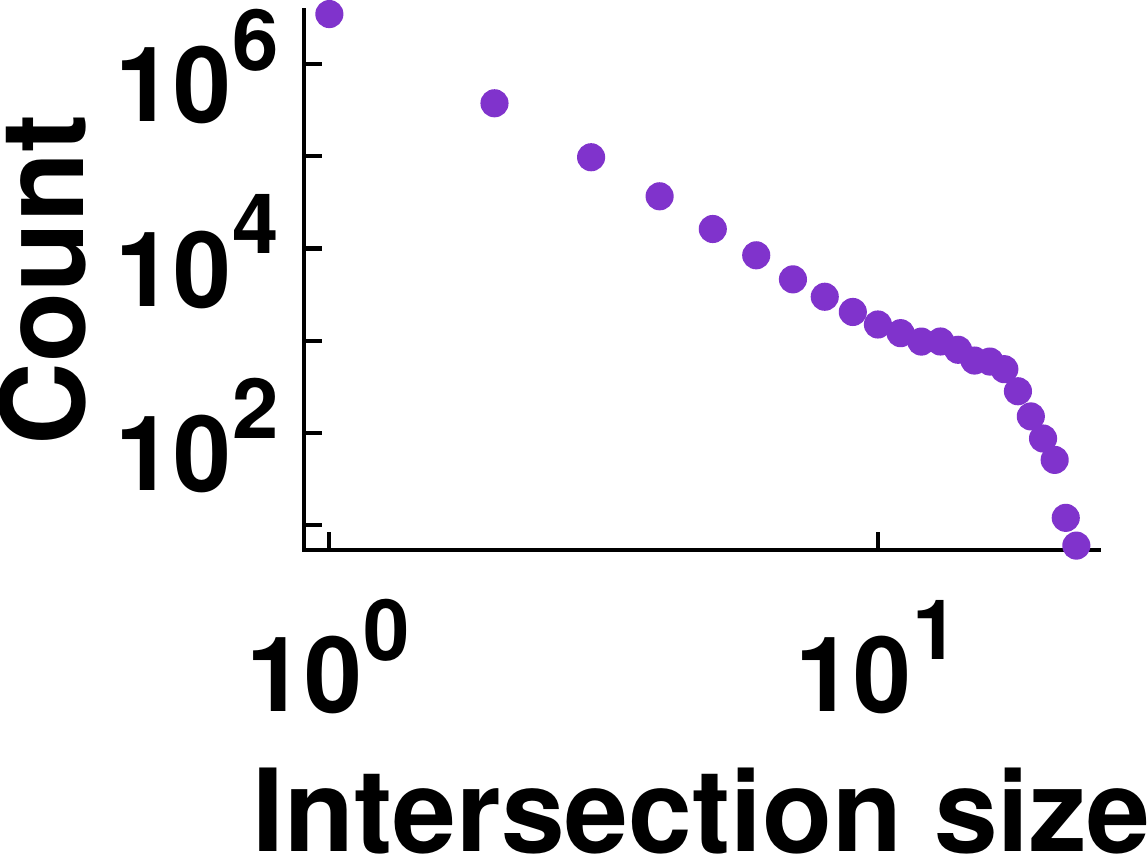} &
			\includegraphics[height=0.762in]{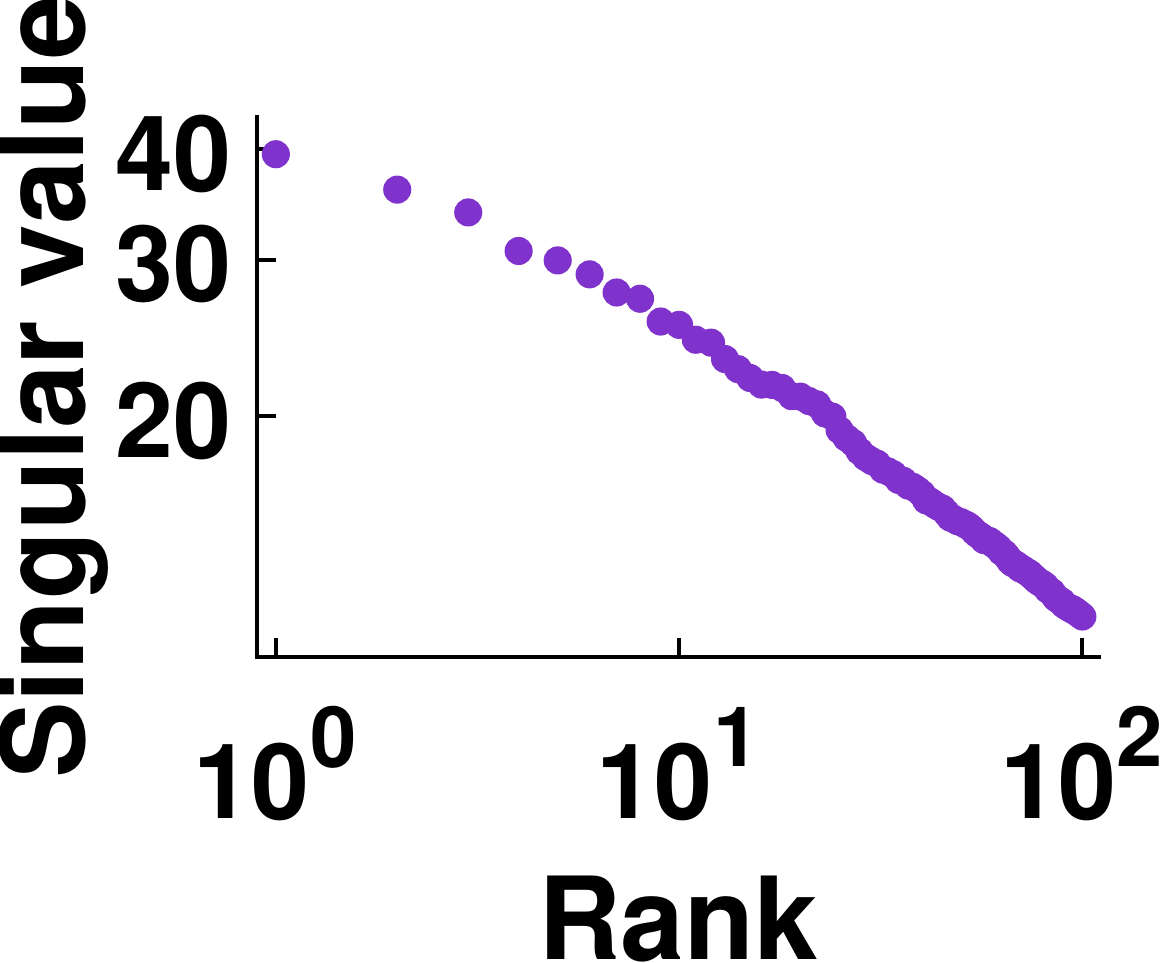} &
			\includegraphics[height=0.762in]{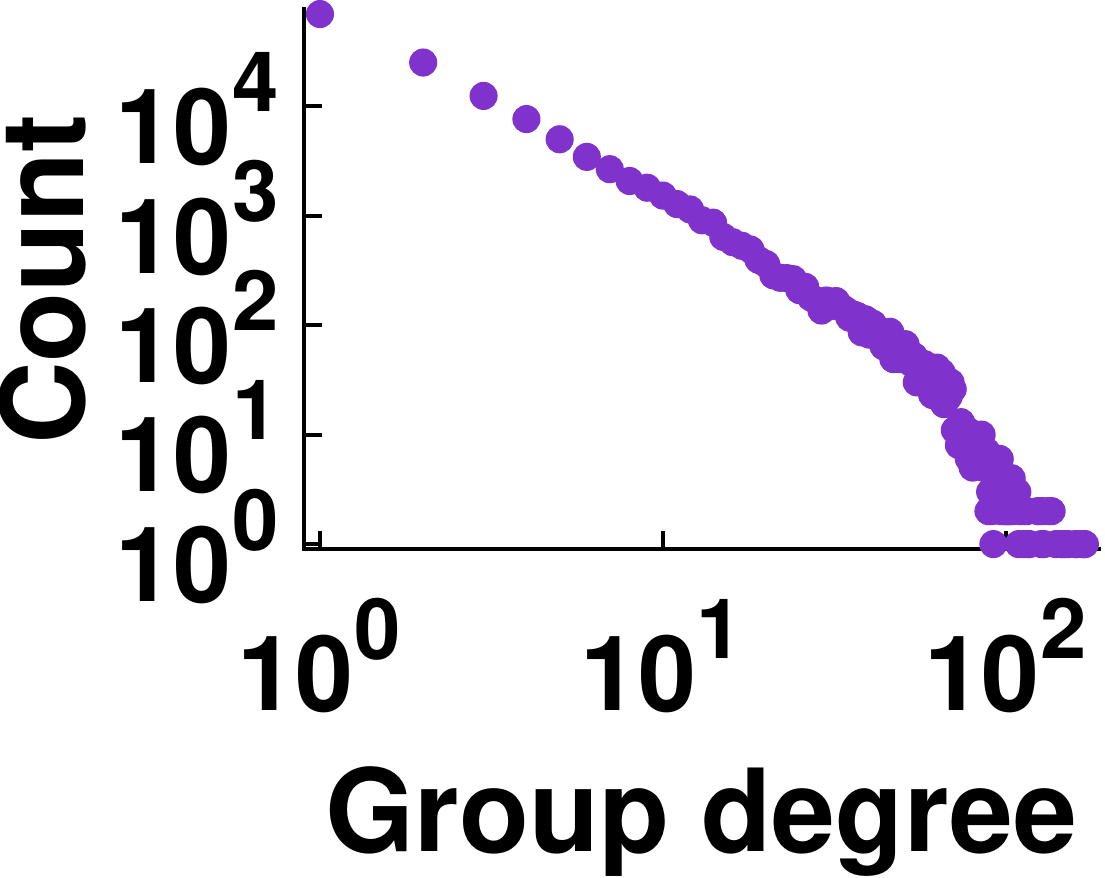} &
			\includegraphics[height=0.762in]{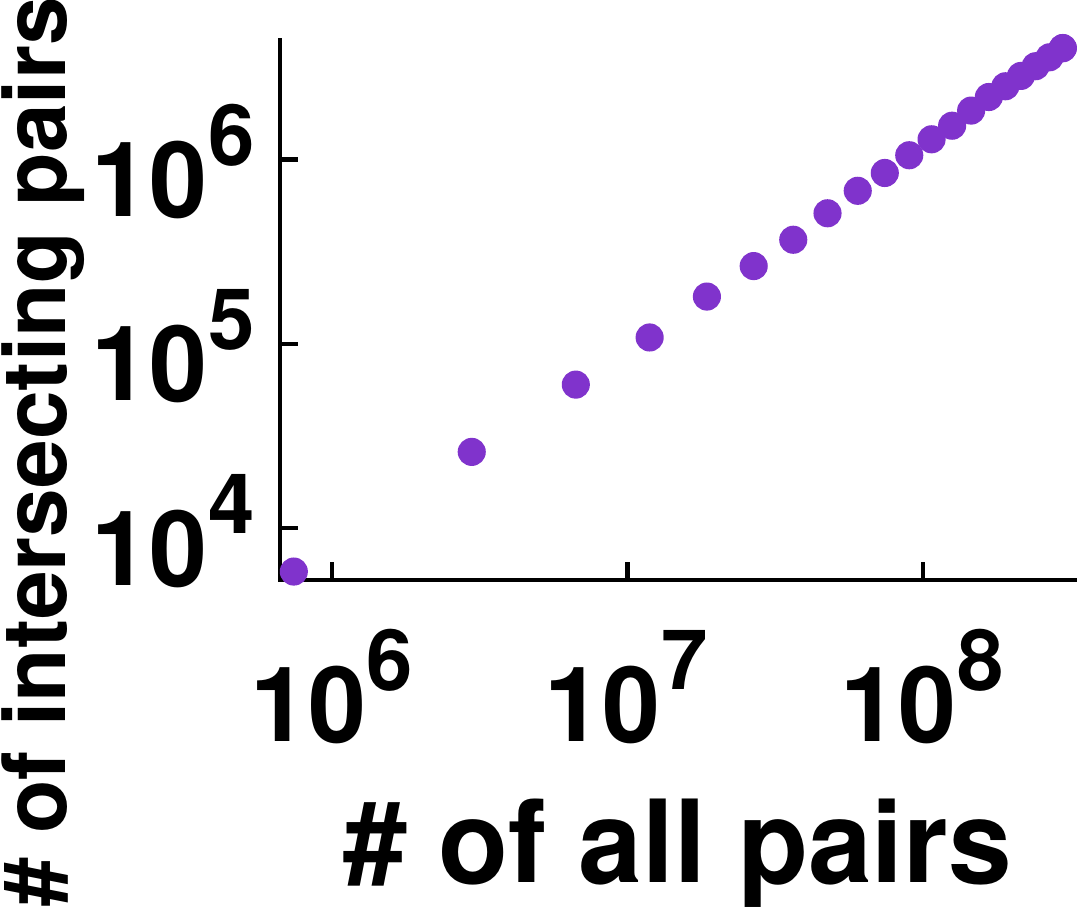} &
			\includegraphics[height=0.762in]{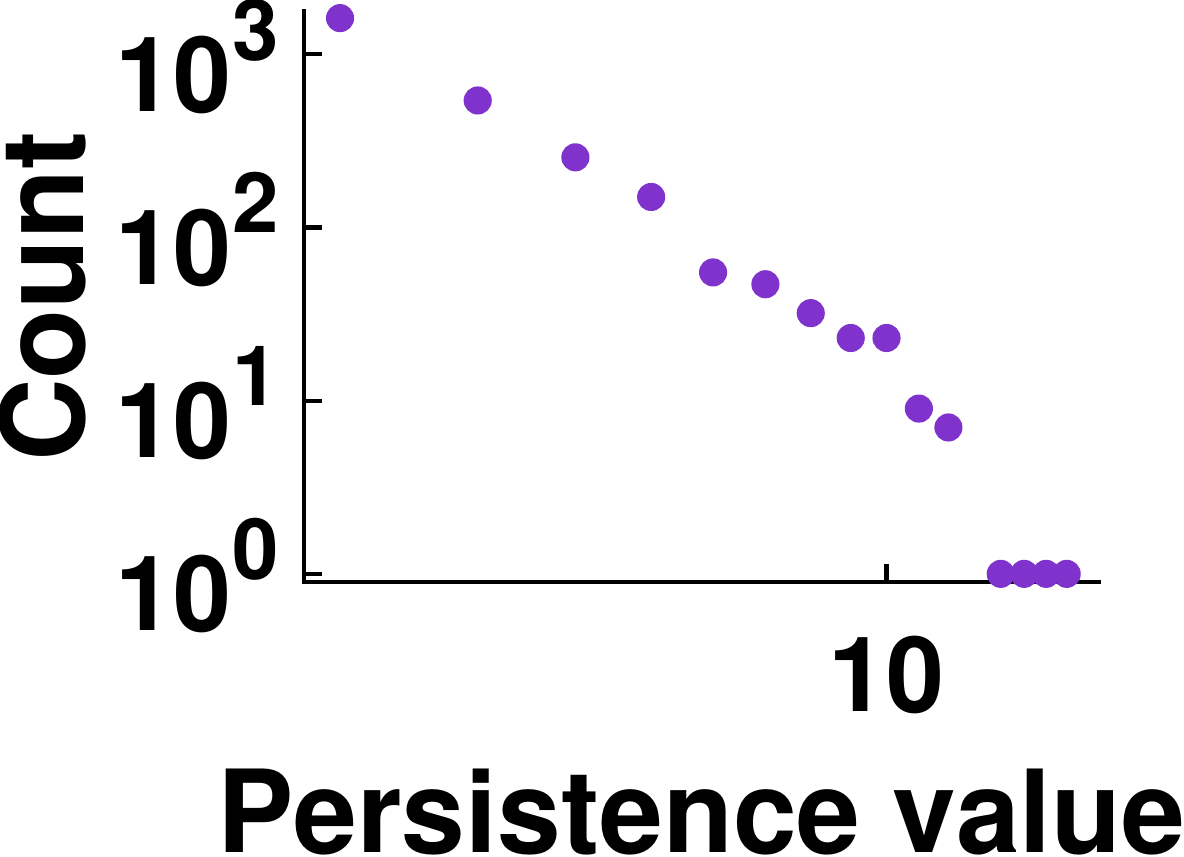}
			\\
			\cmidrule(lr){2-5}
			\cmidrule(lr){6-8}
			\rotatebox[origin=l]{90}{\small \textbf{HyperLLM}}
			\rotatebox[origin=l]{90}{\small \textbf{(Proposed)}}
			& 
			\includegraphics[height=0.762in]{figures/Dfig/Main-1} &
			\includegraphics[height=0.762in]{figures/Dfig/Main-2} &  	 	 			
			\includegraphics[height=0.762in]{figures/Dfig/Main-3} &
			\includegraphics[height=0.762in]{figures/Dfig/Main-4} &
			\includegraphics[height=0.762in]{figures/Dfig/Main-6} &
			\includegraphics[height=0.762in]{figures/Dfig/Main-5} &
			\includegraphics[height=0.762in]{figures/Dfig/Main-8} \\
			\bottomrule
		\end{tabular}
	}
\end{figure*}

\input{Body/2.tex}  

\input{Body/3.tex}  

\begin{figure*}
  \centering
  \includegraphics[width=\textwidth]{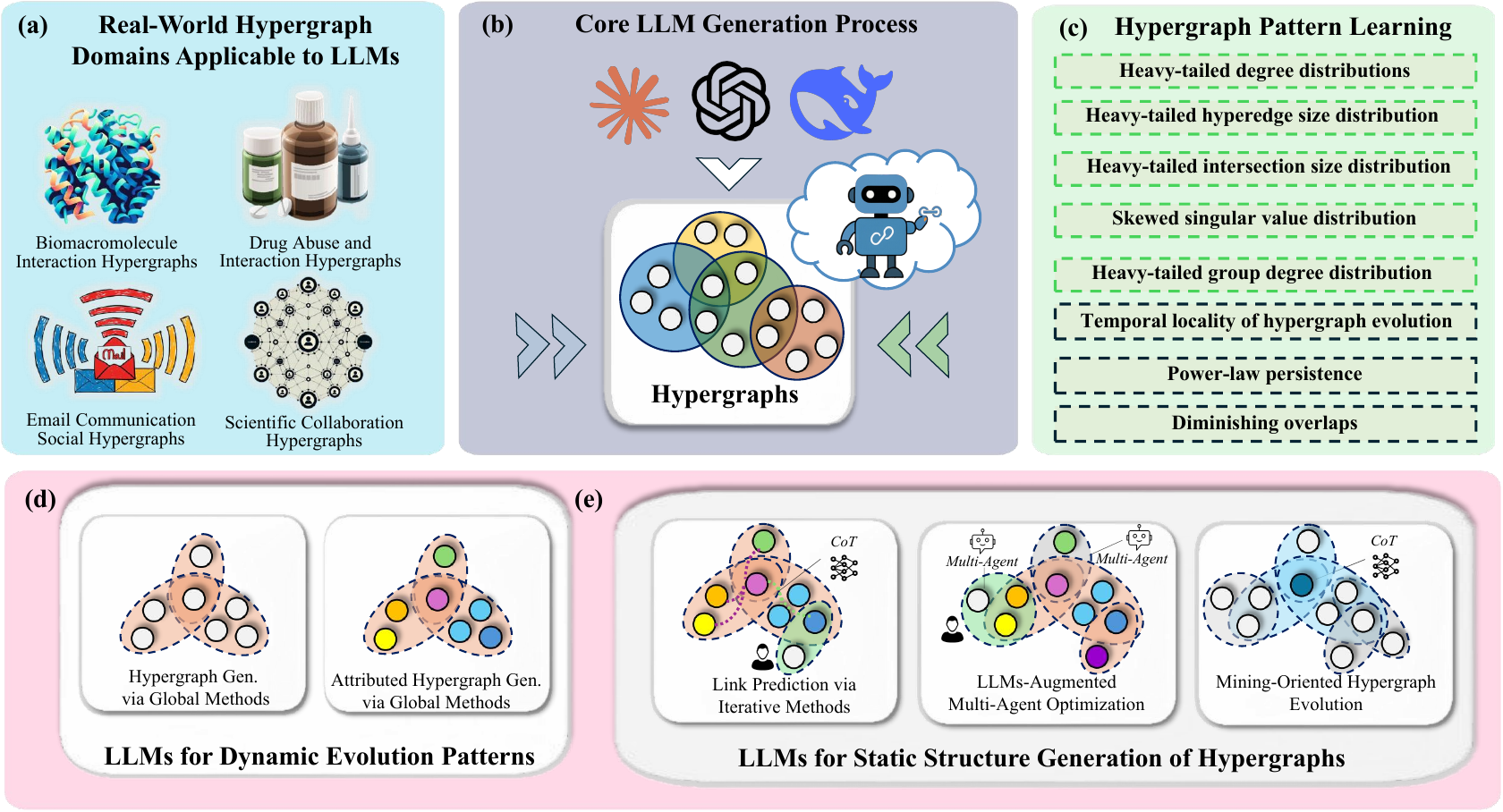}
  \caption{The schematic diagram of the research motivation of this article.}
  \Description{The schematic diagram of the research motivation of this article.}
\end{figure*}

\input{Body/4.tex}  

\begin{figure*}[t]
  \centering
  \includegraphics[width=\textwidth]{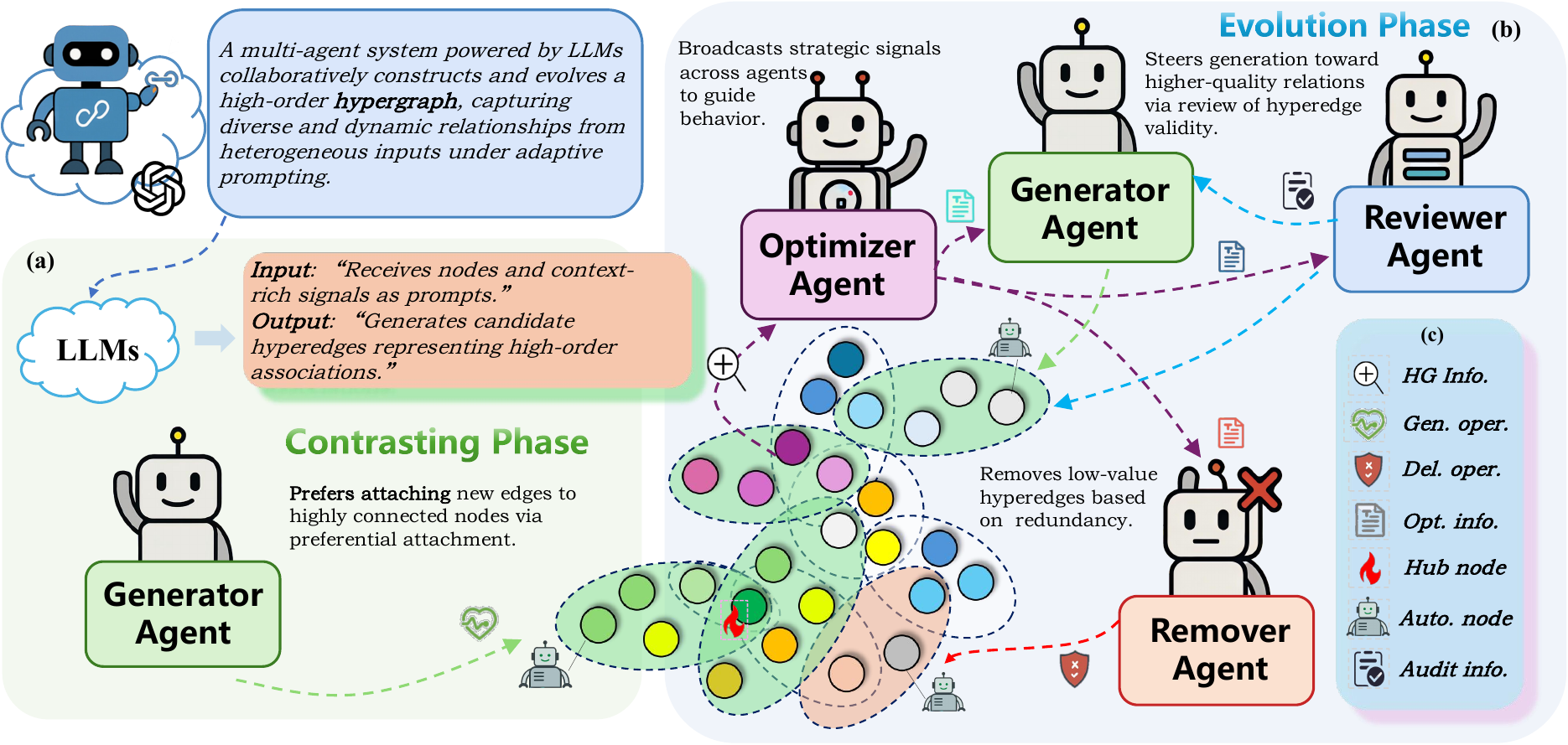}
  \caption{The core mechanism of multi-agent collaborative hypergraph generation.}
  \Description{The core mechanism of multi-agent collaborative hypergraph generation.}
  
  \label{fig:multi_agent}
\end{figure*}

\input{Body/5.tex}  

\input{Body/6.tex}  

\input{Body/7.tex}  

\bibliographystyle{ACM-Reference-Format}
\bibliography{ref}



\end{document}

%% file: Body/1.tex
\section{Introduction}

Networks, as fundamental structures for modeling relationships, are extensively used across diverse domains. To represent high-order interactions involving multiple entities simultaneously, hypergraphs provide a powerful generalization beyond simple pairwise links. Building on this, significant research has been conducted in areas like hypergraph neural networks \cite{ji2020dual, zhao2023dynamic, Kim2024HNN} and higher-order clustering \cite{Carranza2020HOC, yin2018higher}, demonstrating the broad applicability of hypergraph-based analysis.

Despite their importance, a critical challenge persists: the scarcity of high-quality, large-scale hypergraph datasets. This data bottleneck severely hampers the training and evaluation of advanced hypergraph learning algorithms and limits a deeper understanding of the high-order interaction patterns. Consequently, the ability to generate synthetic hypergraphs that faithfully capture the properties of real-world networks is of paramount importance.

Existing hypergraph generation models span several paradigms. These include statistical approaches like configuration models \cite{chodrow2020configuration}, evolutionary models based on preferential attachment \cite{giroire2022preferential, do2020structural}, and algorithmic methods that capture structural features like core periphery organization \cite{Papachristou2022CoreperipheryMF}. Other works have focused on reproducing specific patterns related to hyperedge overlaps \cite{lee2021hyperedges} or temporal dynamics \cite{kook2020evolution, sun2025deep}.

However, these traditional methods suffer from significant limitations. As summarized in Table~\ref{tab:model_comparison}, they often fail to capture the full spectrum of structural and semantic properties of real-world hypergraphs, struggle with computational efficiency, and lack node autonomy, treating entities as passive objects rather than active agents. The advent of Large Language Models (LLMs) offers a promising new direction to overcome these shortcomings.

\begin{table*}[!t]
\centering
\caption{Comparison of Hypergraph Generation Models. The "Validated Criteria" column indicates which of the 25 patterns for criteria (C1-C25) from the survey \cite{lee2025surveyhypergraphminingpatterns}. Our model's validated criteria are linked to their definitions in Section 4.}
\label{tab:model_comparison}
\small 
\setlength{\tabcolsep}{3.5pt} 
\begin{tabular*}{\textwidth}{@{\extracolsep{\fill}}l|cccccc|l@{}}
\toprule
\textbf{Model} & \textbf{Fast} & \textbf{No Large-scale} & \textbf{Node} & \textbf{Inter-} & \textbf{Multi-agent} & \textbf{No Scale} & \textbf{Validated} \\
 & \textbf{Generation} & \textbf{Training} & \textbf{Autonomy} & \textbf{pretability} & \textbf{Collaboration} & \textbf{Limitation} & \textbf{Criteria (C1-C27)} \\
\midrule
CIGAM\cite{Papachristou2022CoreperipheryMF} & \cellcolor{lightpink}{\color{checkgreen}\Checkmark} & \cellcolor{lightpink}{\color{checkgreen}\Checkmark} & {\color{red}$\times$} & \cellcolor{lightpink}{\color{checkgreen}\Checkmark} & {\color{red}$\times$} & {\color{red}$\times$} & C1, C3 \\
Hyper-dK\cite{nakajima2021randomizing} & \cellcolor{lightpink}{\color{checkgreen}\Checkmark} & \cellcolor{lightpink}{\color{checkgreen}\Checkmark} & {\color{red}$\times$} & \cellcolor{lightpink}{\color{checkgreen}\Checkmark} & {\color{red}$\times$} & {\color{red}$\times$} & C1, C4, C7 \\
HyperPA\cite{do2020structural} & {\color{red}$\times$} & \cellcolor{lightpink}{\color{checkgreen}\Checkmark} & {\color{red}$\times$} & {\color{red}$\times$} & {\color{red}$\times$} & \cellcolor{lightpink}{\color{checkgreen}\Checkmark} & C1, C4, C7, C10, C15 \\
HMPA\cite{giroire2022preferential} & {\color{red}$\times$} & \cellcolor{lightpink}{\color{checkgreen}\Checkmark} & {\color{red}$\times$} & \cellcolor{lightpink}{\color{checkgreen}\Checkmark} & {\color{red}$\times$} & \cellcolor{lightpink}{\color{checkgreen}\Checkmark} & C12 \\
HyperFF\cite{kook2020evolution} & \cellcolor{lightpink}{\color{checkgreen}\Checkmark} & \cellcolor{lightpink}{\color{checkgreen}\Checkmark} & {\color{red}$\times$} & {\color{red}$\times$} & {\color{red}$\times$} & \cellcolor{lightpink}{\color{checkgreen}\Checkmark} & C1, C4, C8, C15, C25, C26, C27 \\
DARH\cite{Gallo2024} & \cellcolor{lightpink}{\color{checkgreen}\Checkmark} & \cellcolor{lightpink}{\color{checkgreen}\Checkmark} & {\color{red}$\times$} & \cellcolor{lightpink}{\color{checkgreen}\Checkmark} & {\color{red}$\times$} & {\color{red}$\times$} & C17 \\
HRW\cite{zhang2023efficiently} & {\color{red}$\times$} & \cellcolor{lightpink}{\color{checkgreen}\Checkmark} & {\color{red}$\times$} & \cellcolor{lightpink}{\color{checkgreen}\Checkmark} & {\color{red}$\times$} & \cellcolor{lightpink}{\color{checkgreen}\Checkmark} & C1, C4 \\
CRU\cite{benson2018sequences} & {\color{red}$\times$} & {\color{red}$\times$} & {\color{red}$\times$} & \cellcolor{lightpink}{\color{checkgreen}\Checkmark} & {\color{red}$\times$} & {\color{red}$\times$} & C7, C16, C17 \\
HyRec\cite{choe2025Kronecker} & \cellcolor{lightpink}{\color{checkgreen}\Checkmark} & \cellcolor{lightpink}{\color{checkgreen}\Checkmark} & {\color{red}$\times$} & \cellcolor{lightpink}{\color{checkgreen}\Checkmark} & {\color{red}$\times$} & \cellcolor{lightpink}{\color{checkgreen}\Checkmark} & C1, C4, C7, C8, C10, C11, C14 \\
\midrule
\rowcolor{darkpink}
\textbf{HyperLLM (Ours)} & {\color{checkgreen}\textbf{\Checkmark}} & {\color{checkgreen}\textbf{\Checkmark}} & {\color{checkgreen}\textbf{\Checkmark}} & {\color{checkgreen}\textbf{\Checkmark}} & {\color{checkgreen}\textbf{\Checkmark}} & {\color{checkgreen}\textbf{\Checkmark}} & \textbf{\hyperref[pattern:degree]{C1}, \hyperref[pattern:size]{C4}, \hyperref[pattern:group_degree]{C7}, \hyperref[pattern:intersection]{C8}, \hyperref[pattern:singular]{C15}, \hyperref[pattern:locality]{C17}, \hyperref[pattern:persistence]{C19}, \hyperref[pattern:overlaps]{C25}} \\
\bottomrule
\end{tabular*}
\end{table*}
LLMs have demonstrated a remarkable capacity to comprehend network structures, from pairwise graphs\cite{guo2023gpt4graph} to hypergraphs\cite{feng2025beyond}, and to simulate nuanced human behavior\cite{aher2023replicate, jia2024llmBehavior}, encompassing agent decision-making\cite{sun2025llm}, cognitive biases\cite{ashery2025emergent, chang2025llms}, and emotional responses\cite{li2023large}. Consequently, these advanced reasoning and simulation abilities provide a powerful foundation for generating structurally sound and semantically rich hypergraphs. LLMs possess unique advantages that can address the limitations of existing models: 1) their semantic understanding enables the generation of meaningful, context-aware relationships beyond mere statistical patterns; 2) their reasoning capabilities allow for the creation of globally coherent network structures; and 3) their capacity for behavioral modeling provides a natural framework for simulating the realistic interactions that form complex social and biological systems. As highlighted in Table~\ref{tab:model_comparison}, our proposed HyperLLM is the only framework designed to satisfy all key criteria for a state-of-the-art generator.

This study pioneers the use of large language models in hypergraph modeling, achieving both efficiency and realism. Our main contributions are threefold:
\begin{itemize}[leftmargin=*]
    \item We utilize 8 universal patterns from real-world hypergraphs as a comprehensive set of evaluation metrics, providing a basis for assessing quality and lending interpretability to our framework.
    \item We propose HyperLLM, a comprehensive hypergraph generator that employs a novel methodology based on structured prompts and multi-agent collaboration to produce hypergraphs that are both realistic and semantically coherent.
    \item We conduct extensive experimental validation demonstrating that HyperLLM achieves excellent performance across multiple critical dimensions, including efficiency, realism and portability.
\end{itemize} 

%% file: Body/2.tex
\section{Related Work}
\label{sec:related_work}

This section reviews literature in three key areas: hypergraph generation methods, the emerging use of LLMs for simulating human behavior, and efforts to apply LLMs to network generation.

\subsection{Hypergraph Generation Methods}

Research in hypergraph generation aims to create synthetic networks that faithfully reproduce the structural and temporal patterns of real-world systems. Existing methods can be broadly categorized into static and dynamic approaches. \cite{lee2025surveyhypergraphminingpatterns}
\textit{Static generators} construct a hypergraph in its entirety based on prescribed global or local properties. These include extensions of classical graph models, such as configuration models that preserve fine-grained local statistics like joint degree distributions \cite{nakajima2021randomizing}, and block models designed to embed community structures \cite{deng2024strong}. 
\textit{Dynamic generators} simulate the evolutionary process of a hypergraph, often by adding nodes and hyperedges incrementally. A prominent example is the extension of preferential attachment to hypergraphs, where new nodes are more likely to connect to well-connected existing nodes, thereby generating heavy-tailed degree distributions \cite{do2020structural}. Other dynamic models are inspired by real-world analogies, such as a forest fire, to determine hyperedge formation \cite{kook2020evolution, sun2025deep}.
A separate line of work focuses on generating representative \textit{sub-hypergraphs} from massive networks via biased sampling techniques \cite{choe2024representative}.
While these models have advanced our ability to replicate structural patterns, they often rely on hand-engineered rules and statistical assumptions, limiting their capacity to capture the underlying semantic drivers of hypergraph formation.

\subsection{LLMs for Human Behavior Simulation}

The emergence of Large Language Models (LLMs) has opened new frontiers in simulating complex human behaviors. Prior work has demonstrated LLMs' abilities to realistically simulate human responses and interactions \cite{aher2023replicate,park2023generative,argyle2023samples}, including nuanced aspects like agent decision-making \cite{sun2025llm}, cognitive biases \cite{ashery2025emergent, chang2025llms}, and emotional responses \cite{li2023large}. More recently, research has studied the dynamics of LLM agent populations, showing how social conventions and consensus can arise from their interactions \cite{ashery2024convention,marzo2024consensus}. However, while these studies often simulate interactions over social networks, they tend to make simplistic assumptions about the underlying network structure or require human involvement to build it \cite{park2022social,chuang2023simulating}. This body of work suggests that if LLMs can successfully model the individual and collective behaviors, they should also be capable of generating the resulting network structures themselves.

\subsection{LLMs for Complex Network Generation}

The application of LLMs to graph-structured data is a nascent but rapidly developing field \cite{li2024surveygraphmeetslarge, jin2024largelanguagemodelsgraphs}. Much of the current research focuses on leveraging LLMs for graph-related tasks, such as reasoning over graph structures in natural language \cite{wang23nlgraph}, performing node classification \cite{chen2024exploring}, and enhancing knowledge graph applications \cite{Pan_2024}. A few contemporaneous works have explored using LLMs for social network generation, but have concentrated on traditional pairwise graphs, analyzing the emergence of properties like scale-free degree distributions \cite{marzo2023scalefree} or fundamental principles like preferential attachment and homophily \cite{papachristou2024network, chang2025llms}. Other studies have investigated the risks of demographic and social bias in LLM-based graph generation, revealing that models tend to significantly overestimate political homophily \cite{chang2025llms}. While some research has begun to investigate whether LLMs can comprehend hypergraphs \cite{feng2025beyond}, the direct generation of complex, realistic hypergraphs using LLMs remains a significant and largely unexplored challenge. Our work seeks to bridge this gap. 

%% file: Body/3.tex
\section{Preliminaries}

This section provides the formal definitions and mathematical notation for hypergraphs used throughout this paper.

A hypergraph is formally defined as a pair $H = (V, E)$, where $V = \{v_1, v_2, \ldots, v_n\}$ is a finite set of vertices, and $E = \{e_1, e_2, \ldots, e_m\}$ is a set of hyperedges. Each hyperedge $e_i \subseteq V$ is a non-empty subset of vertices, allowing it to connect any number of nodes. The number of vertices $n = |V|$ is the \textbf{order} of the hypergraph, and the number of hyperedges $m = |E|$ is its \textbf{size}.

The \textbf{cardinality} of a hyperedge $e_i$, denoted $|e_i|$, is the number of vertices it contains. A hypergraph is said to be $k$-uniform if all its hyperedges have the same cardinality $k$. The \textbf{degree} of a vertex $v \in V$ is the number of hyperedges that contain it, defined as:
\begin{equation}
d(v) = \sum_{e \in E} \mathbb{I}(v \in e)
\end{equation}
where $\mathbb{I}(\cdot)$ is the indicator function.

The structure of a hypergraph can be represented by an $n \times m$ \textbf{incidence matrix} $\mathbf{M}$, where its entries are defined as:
\begin{equation}
\mathbf{M}_{ij} = 
\begin{cases} 
1 & \text{if } v_i \in e_j \\
0 & \text{otherwise} 
\end{cases}
\end{equation}
This matrix provides a complete, unambiguous representation of the relationships between vertices and hyperedges. Higher-order relationships can also be represented by an adjacency tensor, which generalizes the concept of an adjacency matrix to capture multi-way interactions, although the incidence matrix is sufficient for the scope of this work. 

%% file: Body/4.tex
\section{Observations and Analysis}

Recent studies have shown that Large Language Models (LLMs) can effectively simulate human behavior, emotions, and social computation. Inspired by this, we investigate whether LLMs can also generate high-quality, realistic hypergraphs. In this section, we identify eight structural and dynamic patterns consistently observed in real-world hypergraphs. We then introduce a mathematical model, inspired by mechanisms in pairwise networks \cite{Pagan2021} but extended to high-order interactions, to explain the "rich get richer" phenomenon. This model not only provides a theoretical foundation for the observed heavy-tailed distributions but also offers a degree of interpretability for the success of LLMs in simulating hypergraph generation.

\subsection{Universal Patterns in Real-World Hypergraphs}
Our analysis of diverse real-world hypergraphs reveals eight universal patterns. Several of these, such as the distributions of degrees, hyperedge sizes, and intersection sizes, are illustrated in Figure~\ref{fig:complete}. These patterns serve as structural and dynamic benchmarks for synthetic hypergraph generation.

\noindent\textbf{P1. Heavy-tailed degree distribution:} \label{pattern:degree} The number of hyperedges connected to a node follows a heavy-tailed distribution, indicating the existence of "hub" nodes that participate in many interactions, a phenomenon commonly described as "the rich get richer."
    
\noindent\textbf{P2. Heavy-tailed hyperedge size distribution:} \label{pattern:size} The number of nodes within hyperedges is also heavy-tailed. While most interactions occur in small groups, a few massive collaborations or events encompass a large number of nodes.

\noindent\textbf{P3. Heavy-tailed intersection size distribution:} \label{pattern:intersection} The number of shared nodes between pairs of hyperedges is heavy-tailed. Most hyperedge pairs have no or very small overlaps, but a few have large intersections, indicating significant structural correlations.\cite{kook2020evolution}

\noindent\textbf{P4. Skewed singular value distribution:} \label{pattern:singular} The singular values of the hypergraph's incidence matrix exhibit a skewed, heavy-tailed distribution. This implies that the hypergraph's structure is low-rank, with a few dominant components capturing most of the structural information. \citep{drineas2004clustering,sarkar2011community,kim2012automated}

\noindent\textbf{P5. Heavy-tailed group degree distribution:} \label{pattern:group_degree} The distribution of community or group sizes is heavy-tailed. This reflects a hierarchical community structure with a few large, overarching communities and many small, tight-knit groups. \citep{benson2018sequences}

\noindent\textbf{P6. Temporal locality of hypergraph evolution:} \label{pattern:locality} The formation of new hyperedges is temporally localized. Nodes that have participated in recent interactions are more likely to participate in future ones, demonstrating short-term memory in the evolutionary process. \citep{benson2018sequences, Gallo2024}

\noindent\textbf{P7. Power-law persistence:} \label{pattern:persistence} The inter-event times of a node's participations in hyperedges follow a power-law distribution. This pattern signifies bursty dynamics, with long inactive periods punctuated by intense flurries of activity. \cite{choo2022persistence}

\noindent\textbf{P8. Diminishing overlaps:} \label{pattern:overlaps} The density of interactions tends to decrease over time. As a hypergraph evolves, newly formed hyperedges are less likely to overlap with existing ones. The density of interactions is defined as:
\begin{equation} \label{eq:doi}
DoI(G_t) := \frac{|\{\{e_i, e_j\} \mid e_i, e_j \in E_t, e_i \cap e_j \neq \emptyset \}|}{|\{\{e_i, e_j\} \mid e_i, e_j \in E_t\}|}
\end{equation}
where $E_t$ is the set of hyperedges at time $t$. The numerator is the number of intersecting pairs of hyperedges, and the denominator is the total number of pairs. \cite{kook2020evolution}

These eight patterns are intuitively or directly connected to our generative mechanism. The proposed mathematical model, a more stringent form of the power-law, provides a unified explanation for these heavy-tailed phenomena.

\subsection{A Microscopic Model for High-Order Preferential Attachment}
To explain the prevalent heavy-tailed distributions, we develop a generative model based on microscopic dynamics. This model extends preferential attachment from pairwise graphs to high-order hypergraphs, incorporating node quality and collaborative inertia.

Consider a hypergraph $H=(V,E)$ with $N=|V|$ nodes. Each node $i$ possesses an intrinsic quality, and we rank them accordingly, with $rank_i \in \{1, ..., N\}$ representing the rank of node $i$ (where 1 is the best). The model also includes a \textbf{collaborative inertia or activation threshold} parameter $\alpha > 0$. When $\alpha=0$, the model simplifies to the classic Zipf law.

At each time step, a new hyperedge of size $k$ is generated. An "initiator" node is chosen. The remaining $k-1$ collaborators are then selected based on a preferential attachment mechanism. The probability of selecting a node $i$ depends on its rank $r_i$ and the inertia $\alpha$. We define an effective reachability probability for collaborator selection:
\begin{equation}
P_{\text{reach}}(i) = \frac{(r_i+\alpha)^{-\gamma}}{\sum_{j=1}^{N} (r_j+\alpha)^{-\gamma}}, \quad \gamma > 0
\end{equation}
Furthermore, we introduce a \textbf{quality filter}, asserting that a hyperedge is only successfully formed if all selected nodes $q_j$ have a quality higher than a certain threshold $q_{th}$. This reflects the fact that high-order collaboration requires a minimum quality commitment from all participants.

Let $d_i(t)$ be the degree of node $v_i$ at time $t$. The probability that node $v_i$ is chosen as a collaborator in a single step is:
\begin{equation}
P_{\text{sel}}(i) = \mathbb{I}[q_i>q_{th}] \cdot \frac{(r_i+\alpha)^{-\gamma}}{\sum_{j:q_j>q_{th}} (r_j+\alpha)^{-\gamma}}
\end{equation}
where $\mathbb{I}[\cdot]$ is the indicator function for the quality filter. If we assume a constant rate of hyperedge arrival $\lambda$, the degree evolution can be described by the continuous-time master equation:
\begin{equation}
\frac{d}{dt} \mathbb{E}[d_i(t)] = \lambda \cdot P_{\text{sel}}(i)
\end{equation}

Assuming the system reaches a steady state after a long time $T$, the expected degree of node $i$ is:
\begin{equation}
d_i = \mathbb{E}[d_i(T)] = \lambda T \cdot P_{\text{sel}}(i) = \lambda T \cdot \frac{(r_i+\alpha)^{-\gamma}}{\sum_{j \in I^*} (r_j+\alpha)^{-\gamma}}
\end{equation}
where $I^*$ is the set of high-quality nodes that pass the filter. This simplifies to the Zipf-Mandelbrot form:
\begin{equation} \label{eq:zipf_mandelbrot}
d_i = A \cdot (r_i+\alpha)^{-\gamma}
\end{equation}
where $A = \frac{\lambda T}{\sum_{j \in I^*} (r_j+\alpha)^{-\gamma}}$ is a normalization constant. For $\gamma=1$, the denominator can be approximated by a Hurwitz zeta function or, for large $N$, by $\ln((N+\alpha)/\alpha)$.

This derivation shows how the Zipf-Mandelbrot law can emerge from microscopic dynamics. This model is not merely abstract; its components have conceptual parallels in our HyperLLM framework. For instance, a node's rank ($r_i$) relates to its perceived influence, the inertia ($\alpha$) is reflected in review thresholds, and the quality filter ($q_i$) is enacted by agents like the \textbf{Reviewer} and \textbf{Remover}. This correspondence suggests our agent-based system succeeds by implicitly simulating such fundamental mechanisms to ensure structural fidelity. 

%% file: Body/5.tex
\section{Hypergraph Generators}
\label{sec:generators}

Our proposed framework, HyperLLM, generates hypergraphs through a process that reflects real-world network development. It consists of two key stages: a Construction Phase and an Evolution Phase. The construction phase uses an efficient iterative algorithm to quickly build the initial network. Then, the evolution phase applies a multi-agent system (MAS), where four specialized LLM-powered agents collaboratively refine and expand the hypergraph, maintaining both structural coherence and semantic depth.

\subsection{High-order Language Prompts}

At the core of HyperLLM are carefully engineered prompts that harness the reasoning capabilities of LLMs. Our framework operates on nodes conceptualized as rich entities, each possessing a unique set of attributes and a persona. The prompts for the four agents in the evolution phase are designed to be generic, allowing HyperLLM to be adapted across various domains, from social and biological networks to scientific collaborations.

\vspace{0.2cm}
\textbf{(1) Generator Agent Template:} This agent is tasked with creating new hyperedges by identifying potential collaborations centered around a specific entity. The prompt guides the LLM to consider both entity attributes and the existing network topology.

\begin{tcolorbox}[top=2pt, bottom=2pt, left=4pt, right=4pt]
\textbf{System prompt:} You are a hypergraph relationship generator. Your task is to form a new collaborative group around a central entity.

\textbf{User content:}
A new collaboration is being formed in a [Domain] network.

  \textbf{Central Entity:} [Entity ID]

  \textbf{Attributes:} [List of Attributes and Values]

  \textbf{Local Context:} [Summary of existing relationships involving the entity]

  \textbf{Task:} Propose a new hyperedge of size [k] that includes the central entity. The group should be semantically coherent and structurally sound based on the context.
\end{tcolorbox}

\vspace{0.3cm}
\textbf{(2) Reviewer Agent Template:} This agent assesses the quality and viability of a proposed hyperedge, acting as a quality control mechanism.

\begin{tcolorbox}[top=2pt, bottom=2pt, left=4pt, right=4pt]
\textbf{System prompt:} You are a hypergraph relationship reviewer. Your task is to validate a candidate hyperedge.

\textbf{User content:}
Please review the following candidate hyperedge:

  \textbf{Candidate Hyperedge:} {[List of Entity IDs]}

  \textbf{Entity Details:} [Summary of attributes for all entities in the candidate hyperedge]

  \textbf{Evaluation Criteria:}
    - Internal Cohesion: Are the members a good fit?
    - Network Impact: How does this group affect the overall structure?
  
    \textbf{Decision:} Output "APPROVE" or "REJECT".
\end{tcolorbox}

\vspace{0.3cm}
\textbf{(3) Remover Agent Template:} This agent prunes the network by identifying and removing stale or low-quality hyperedges, ensuring the long-term health of the hypergraph.

\begin{tcolorbox}[top=2pt, bottom=2pt, left=4pt, right=4pt]
\textbf{System prompt:} You are a hypergraph network curator. Your task is to identify and remove redundant or low-quality hyperedges.

\textbf{User content:}
Analyze the provided list of hyperedges.

  \textbf{Hyperedges:} [List of existing hyperedges]

  \textbf{Global Strategy:} [Current network-wide optimization goal, e.g., "ENHANCE\_DIVERSITY"]
  
  \textbf{Task:} Identify indices of hyperedges that are redundant, internally incoherent, or conflict with the global strategy. Output indices or "NONE".
\end{tcolorbox}

\vspace{0.3cm}
\textbf{(4) Optimizer Agent Template:} This agent performs a global assessment of the evolving hypergraph and provides strategic guidance for other agents during each evolution step. Its decisions are informed by general network statistics and diversity measures rather than domain-specific rules. Based on the current state of the network, the optimizer may, for example, recommend increasing connections, enhancing diversity, reducing clustering, or simply maintaining the existing structure. These strategies are illustrative rather than prescriptive, ensuring the approach remains adaptable across different contexts and data sources.

\begin{tcolorbox}[top=2pt, bottom=2pt, left=4pt, right=4pt]
\textbf{System prompt:} You are a network strategy analyst. Your task is to assess the entire hypergraph and provide a strategic directive for the next evolution step.

\textbf{User content:}
Analyze the current hypergraph state.

  \textbf{Network Statistics:} [e.g., density, clustering, component sizes, diversity metrics]

  \textbf{Task:} Based on the analysis, choose one strategic directive that will most effectively improve the network's quality and realism.
  
  \textbf{Decision:} Output corresponding optimization suggestions.
\end{tcolorbox}

\subsection{Generation Algorithms}

HyperLLM employs two complementary algorithms. The first, \textbf{Iterative Local Generation}, is used during the construction phase for rapid, efficient network building. The second, \textbf{Multi-Agent Collaborative Generation}, orchestrates the four specialized agents during the evolution phase for fine-grained, high-quality network refinement.

\begin{algorithm}[H]
\caption{Iterative Local Generation}
\label{alg:iterative_local}
\begin{algorithmic}[1]
\REQUIRE Set of entities $V$, Target number of hyperedges $M$
\ENSURE A foundational hypergraph $H_0 = (V, E_0)$

\STATE $E_0 \leftarrow \emptyset$
\FOR{$i = 1$ to $M$}
    \STATE $v_{c} \leftarrow \textsc{SelectEntity}(V)$ \COMMENT{Select a central entity}
    \STATE $k \leftarrow \textsc{DetermineHyperedgeSize}()$
    \STATE $C_{local} \leftarrow \textsc{GetLocalContext}(v_c, E_0)$
    
    \STATE $P \leftarrow \textsc{BuildGeneratorPrompt}(v_c, C_{local}, k)$
    \STATE $E_{cand} \leftarrow \textsc{LLMQuery}(P)$
    
    \IF{\textsc{Validate}( $E_{cand}$ )}
        \STATE $E_0 \leftarrow E_0 \cup \{E_{cand}\}$
    \ENDIF
\ENDFOR
\RETURN $H_0 = (V, E_0)$
\end{algorithmic}
\end{algorithm}

The multi-agent algorithm forms the core of the evolution phase. In each step, the \textbf{Optimizer} agent first assesses the global state of the network and broadcasts a strategic directive. This directive acts as a communication signal that coordinates the actions of the other agents. The \textbf{Remover} may then prune edges that conflict with the strategy, while the \textbf{Generator} and \textbf{Reviewer} work in tandem to create new, high-quality edges that align with the global goal, as detailed in Algorithm \ref{alg:multi_agent_collaborative}. This creates a feedback loop where global network health guides local modifications.

\begin{algorithm}[H]
\caption{Multi-Agent Collaborative Generation}
\label{alg:multi_agent_collaborative}
\begin{algorithmic}[1]
\REQUIRE Initial hypergraph $H^{t-1}=(V, E^{t-1})$, Agent set $\mathcal{A}$
\ENSURE Evolved hypergraph $H^t=(V, E^t)$

\STATE // 1. Global Strategy Formulation
\STATE $\mathcal{S}_t \leftarrow \textsc{Optimizer.Analyze}(H^{t-1})$
\COMMENT{Optimizer sets global strategy}

\STATE // 2. Network Pruning
\STATE $P_{rem} \leftarrow \textsc{BuildRemoverPrompt}(E^{t-1}, \mathcal{S}_t)$
\STATE $E_{rem} \leftarrow \textsc{Remover.Query}(P_{rem})$
\STATE $E' \leftarrow E^{t-1} \setminus E_{rem}$

\STATE // 3. Candidate Generation
\STATE $E_{new} \leftarrow \emptyset$
\FOR{each generation attempt}
    \STATE $v_c \leftarrow \textsc{SelectEntity}(V)$
    \STATE $P_{gen} \leftarrow \textsc{BuildGeneratorPrompt}(v_c, E', \mathcal{S}_t)$
    \STATE $e_{cand} \leftarrow \textsc{Generator.Query}(P_{gen})$
    
    \STATE // 4. Collaborative Review
    \STATE $P_{rev} \leftarrow \textsc{BuildReviewerPrompt}(e_{cand}, \mathcal{S}_t)$
    \IF{\textsc{Reviewer.Query}( $P_{rev}$ ) = \textsc{APPROVE}}
        \STATE $E_{new} \leftarrow E_{new} \cup \{e_{cand}\}$
    \ENDIF
\ENDFOR

\STATE $E^t \leftarrow E' \cup E_{new}$
\RETURN $H^t = (V, E^t)$
\end{algorithmic}
\end{algorithm} 

%% file: Body/6.tex
\section{Experiments}

\begin{table*}[t!]
	\caption{\textbf{HyperLLM Fits Real-World Hypergraphs.} Comparison of HyperLLM against baselines on fitting structural patterns across 8 datasets. The question mark (?) indicates that size distribution is not strictly required; for small-scale datasets, our model can achieve high-fidelity generation even without this input, showcasing its flexibility. The \textcolor{blue}{best}, \textcolor{darkgreen}{second-best}, and \textcolor{darkyellow}{third-best} average rankings are highlighted.}
	\label{tab:llm_fitting}
	\centering
	\footnotesize
	\begin{tabular}{l|cccc||cccccccc|c}
	  \toprule
	  & \multicolumn{4}{c||}{\textbf{Input Information}} & \multicolumn{8}{c|}{\textbf{Avg. Ranking (8 Datasets)}} & \\
	  & \textbf{NN} & \textbf{DD} & \textbf{SD} & \textbf{OI} & \textbf{Deg. Dist.} & \textbf{Size Dist.} & \textbf{Group Deg.} & \textbf{Inter. Dist.} & \textbf{Sing. Val.} & \textbf{Dimin. Ovlp.} & \textbf{Persistence} & \textbf{Temp. Loc.} & \textbf{Avg.} \\
	  \midrule
	  HyperCL & \Checkmark & \Checkmark & \Checkmark & & \textcolor{blue}{1.88} & \textcolor{blue}{1.00} & \textcolor{blue}{1.63} & 4.38 & \textcolor{darkyellow}{3.00} & 5.50 & 5.50 & 6.00 & \textcolor{darkyellow}{2.92} \\
	  HyperFF & \Checkmark & & & & 4.38 & 5.25 & 4.25 & 5.13 & 5.38 & 3.38 & \textcolor{darkyellow}{2.50} & \textcolor{darkyellow}{2.75} & 3.70 \\
	  HyperPA & \Checkmark & & \Checkmark & \Checkmark & \textcolor{darkyellow}{3.00} & 4.50 & \textcolor{darkgreen}{2.75} & \textcolor{darkyellow}{2.88} & 4.88 & 5.13 & \textcolor{darkgreen}{2.38} & \textcolor{darkgreen}{2.50} & \textcolor{darkgreen}{2.86} \\
	  HyperLAP & \Checkmark & \Checkmark & \Checkmark & \Checkmark & 4.50 & \textcolor{blue}{1.00} & 4.50 & \textcolor{blue}{1.88} & 3.13 & \textcolor{darkgreen}{2.38} & 5.50 & 6.00 & 3.31 \\
	  HyRec & \Checkmark & & & \Checkmark & 4.63 & 5.25 & 4.38 & 4.00 & \textcolor{blue}{2.00} & \textcolor{blue}{1.75} & 3.00 & 3.38 & 3.33 \\
	  \midrule
	  \textbf{HyperLLM} & \Checkmark & & (?) & & \textcolor{darkgreen}{2.63} & \textcolor{darkgreen}{3.00} & \textcolor{darkyellow}{3.50} & \textcolor{darkgreen}{2.00} & \textcolor{darkgreen}{2.38} & \textcolor{darkyellow}{2.88} & \textcolor{blue}{\textbf{2.13}} & \textcolor{blue}{\textbf{1.38}} & \textcolor{blue}{\textbf{2.13}} \\
	   \bottomrule
	 \end{tabular}
\end{table*}

\begin{figure*}[t!]
	\centering
	\caption{
		\textbf{Comparison between a real hypergraph and those generated by HyperLLM with different attachment probabilities.} The results, particularly for preferential attachment probability 0.85, closely mimic the real data patterns.
	}
	\label{fig:model_params}
	\scalebox{0.82}{
		\begin{tabular}{c|ccccccc}
			\toprule
			\multicolumn{1}{c}{} &
			\multicolumn{5}{c}{\textsc{Structural Patterns}} &
			\multicolumn{2}{c}{\textsc{Dynamic Patterns}}	
			\\
			\cmidrule(lr){2-6}
			\cmidrule(lr){7-8}
			\multicolumn{1}{c}{} &
			Degrees &
			Hyperedge Sizes &
			Intersection Sizes &
			Singular Values &
			Group Degree &
      		Intersecting Pairs &
			Power-law persistence
			\\
			\cmidrule(lr){2-6}
			\cmidrule(lr){7-8}
			\rotatebox[origin=l]{90}{(P = 0.55)} 
			
			&
			\includegraphics[height=0.762in]{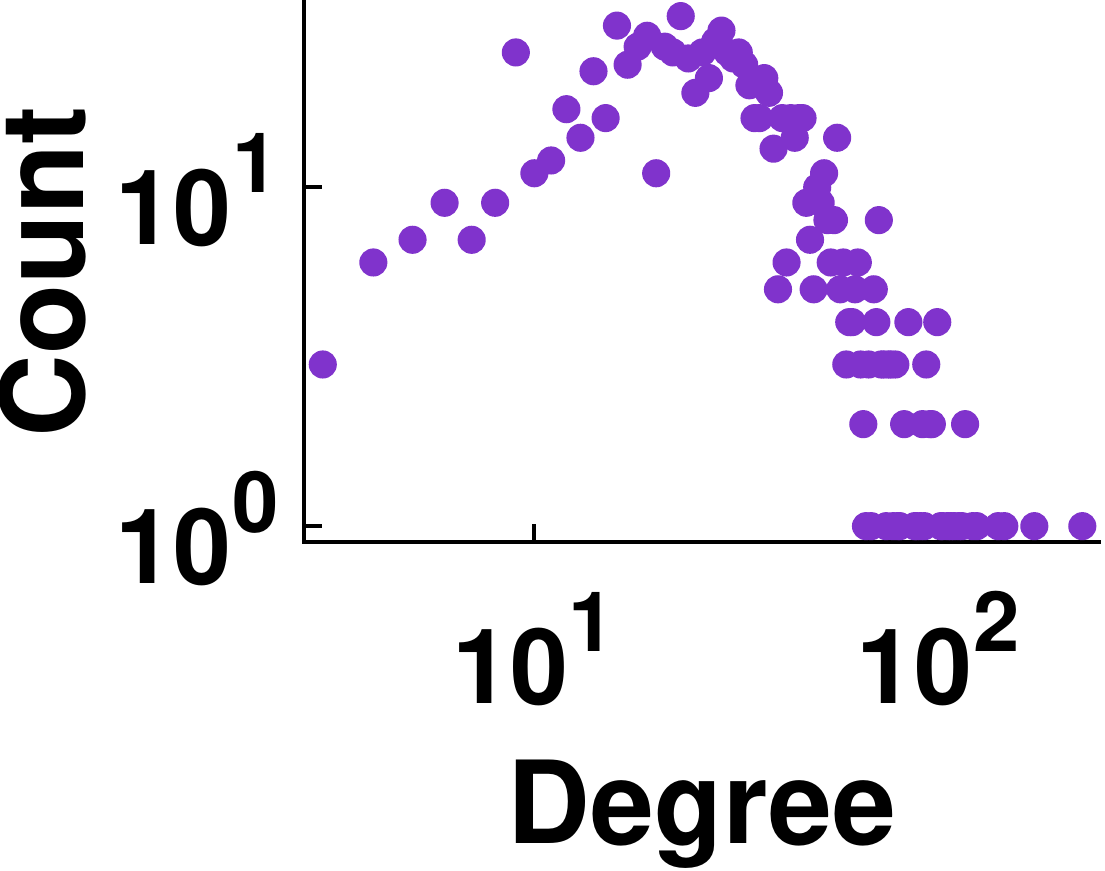} &
			\includegraphics[height=0.762in]{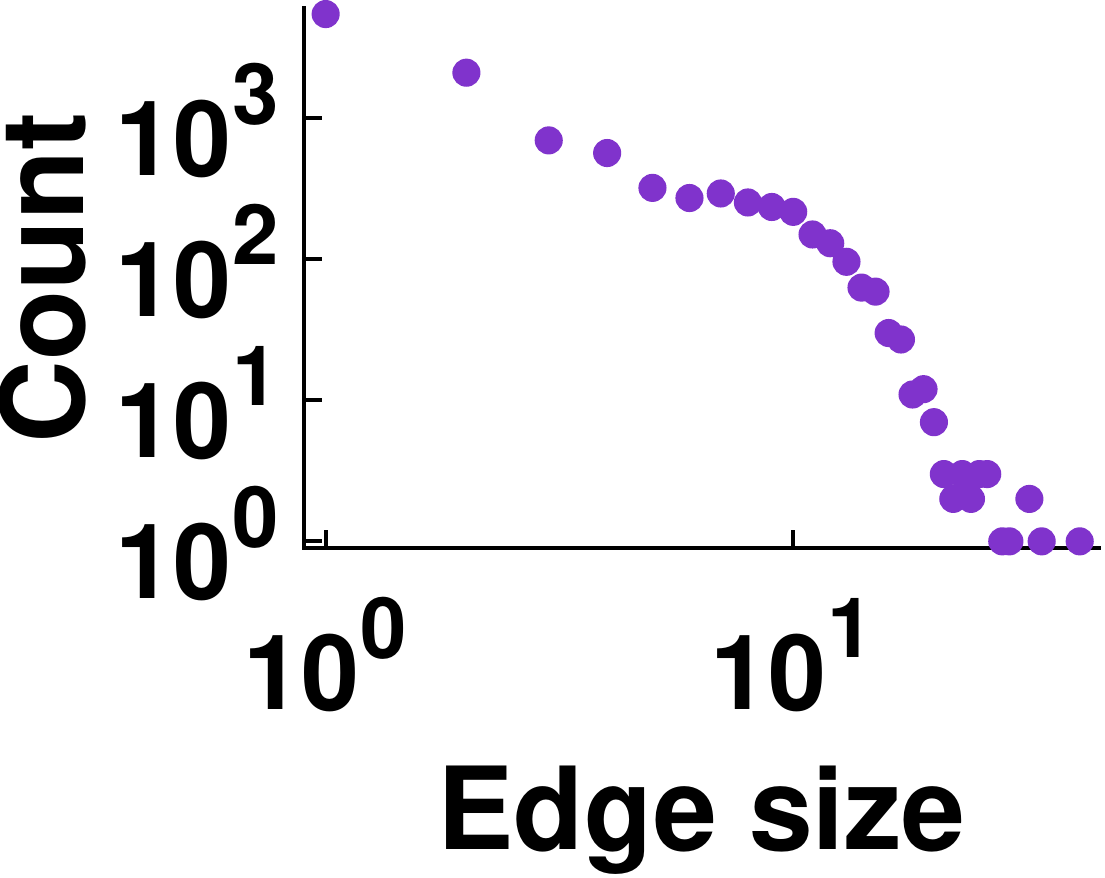} &
			\includegraphics[height=0.762in]{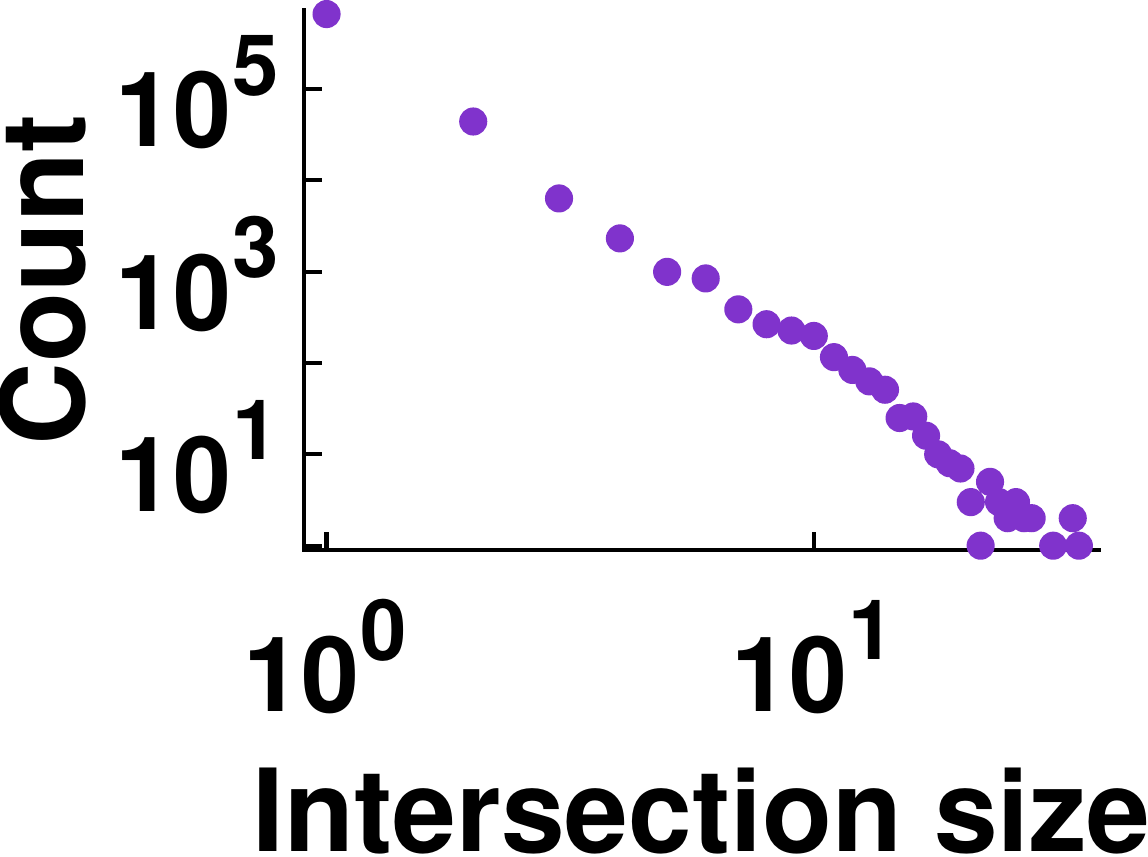} &
			\includegraphics[height=0.762in]{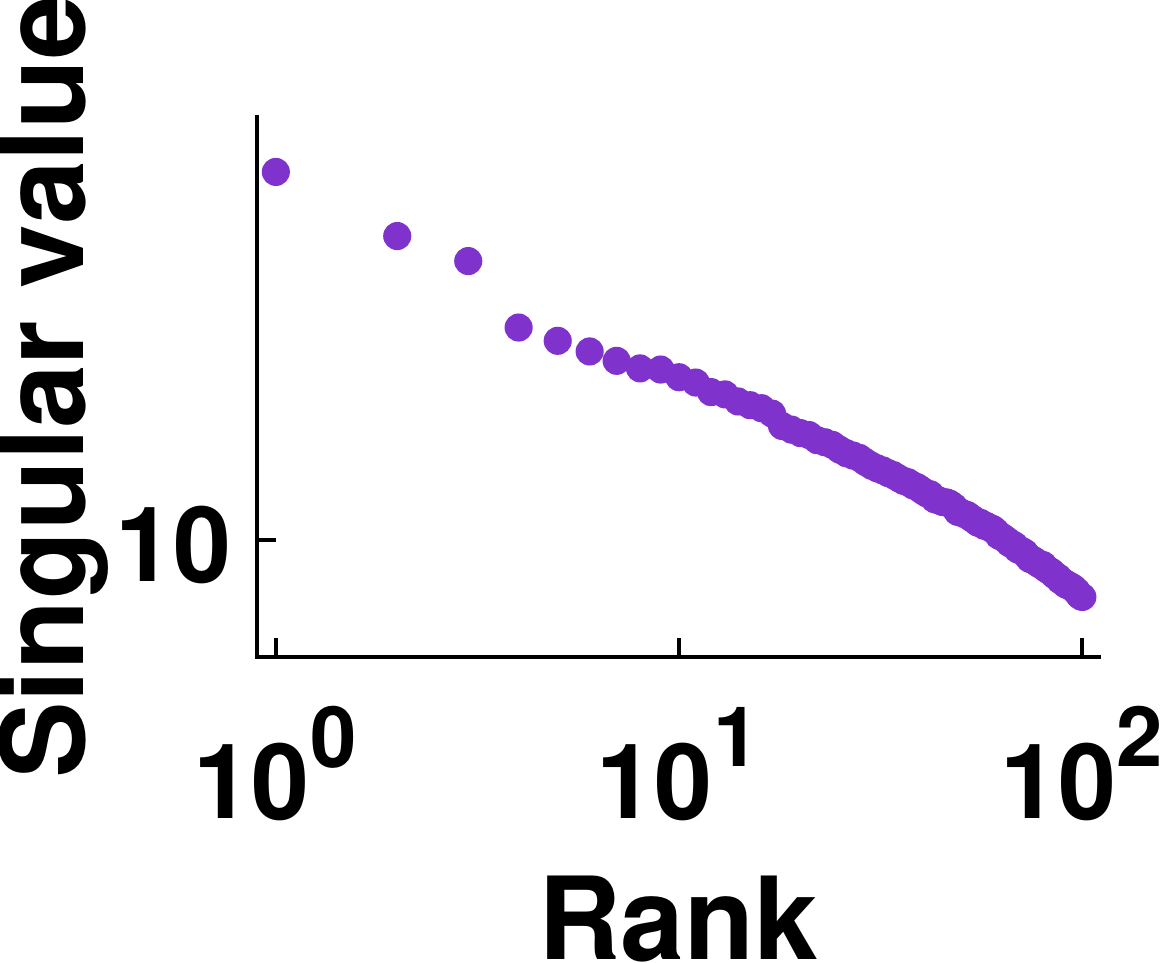} &
			\includegraphics[height=0.762in]{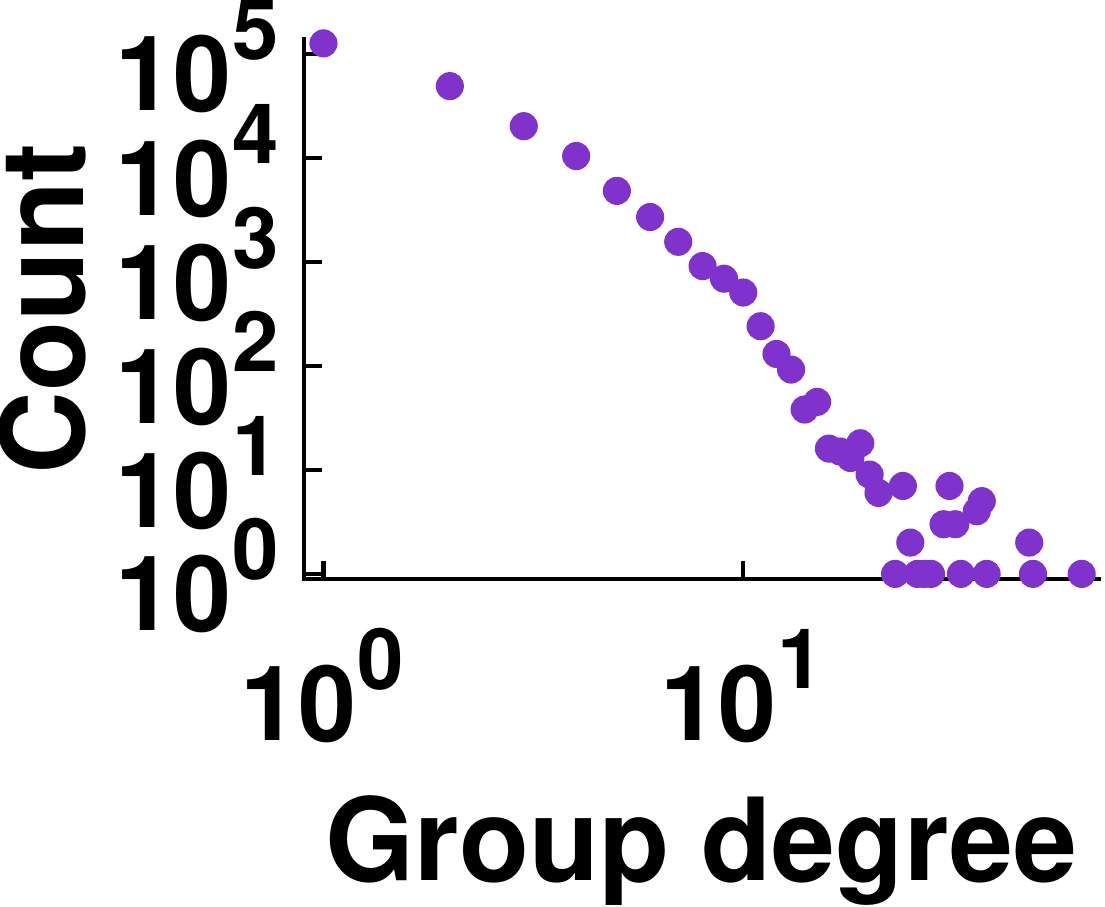} &
			\includegraphics[height=0.762in]{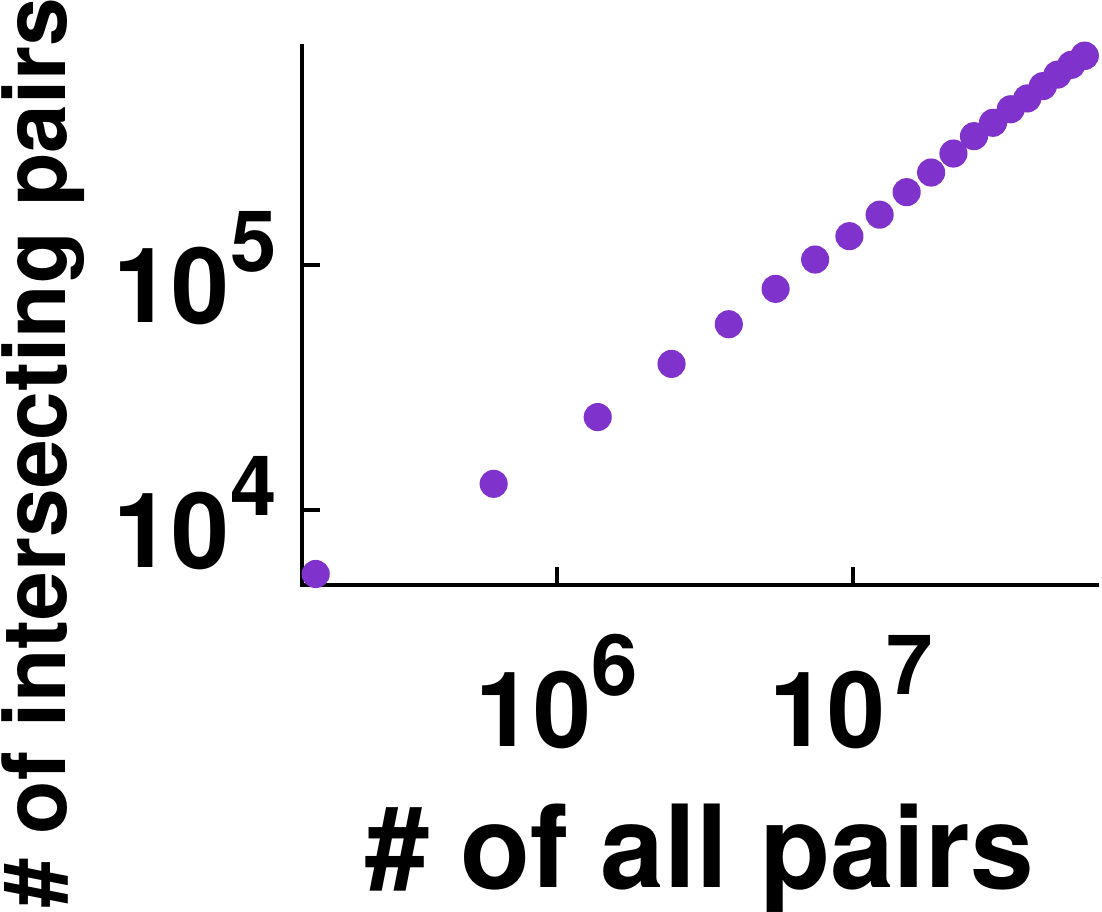} &
			\includegraphics[height=0.762in]{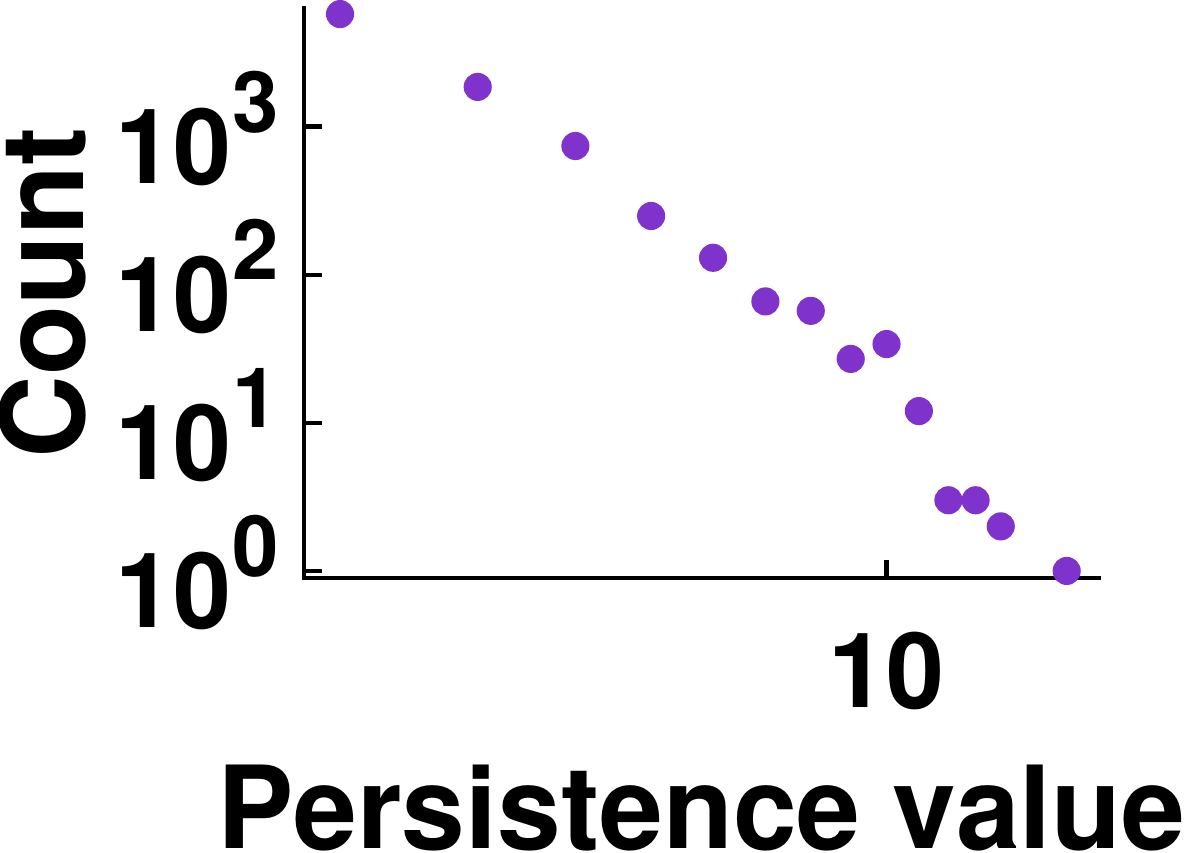}
			\\

			\rotatebox[origin=l]{90}{(P. = 0.65)}
			&
			\includegraphics[height=0.762in]{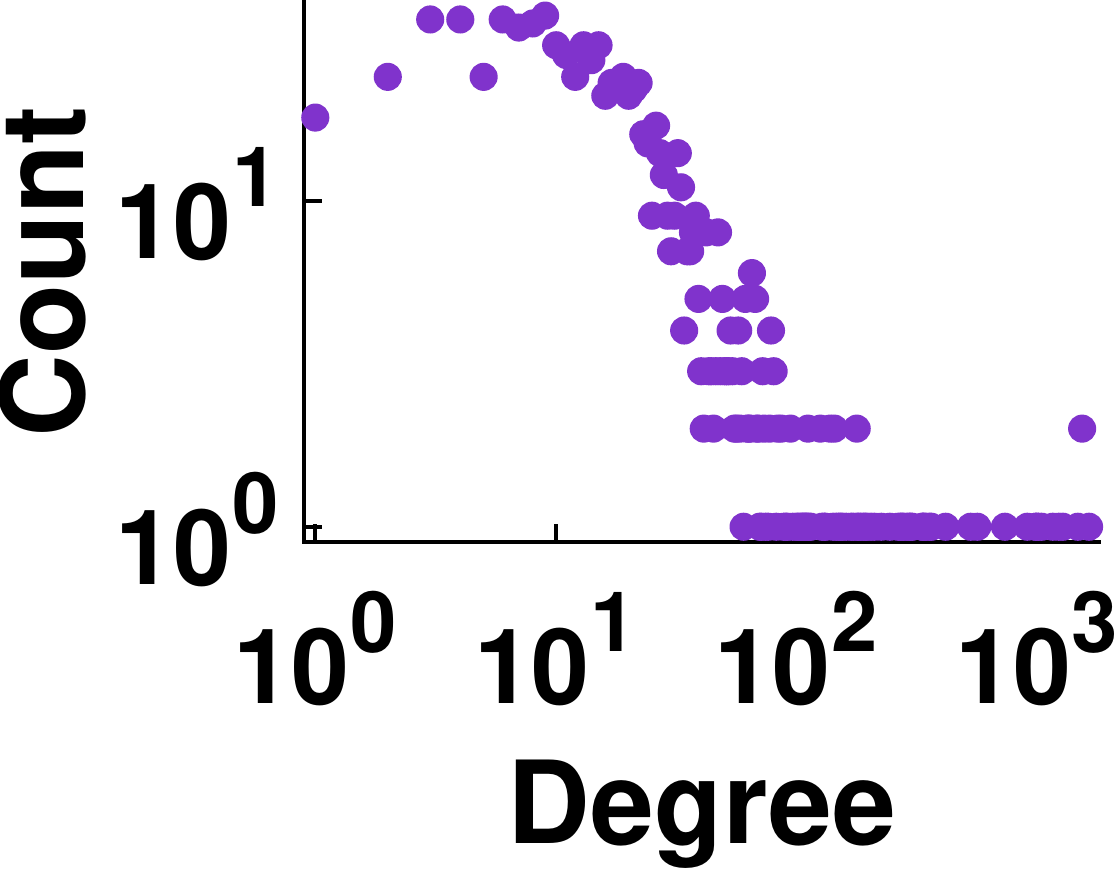} &
			\includegraphics[height=0.762in]{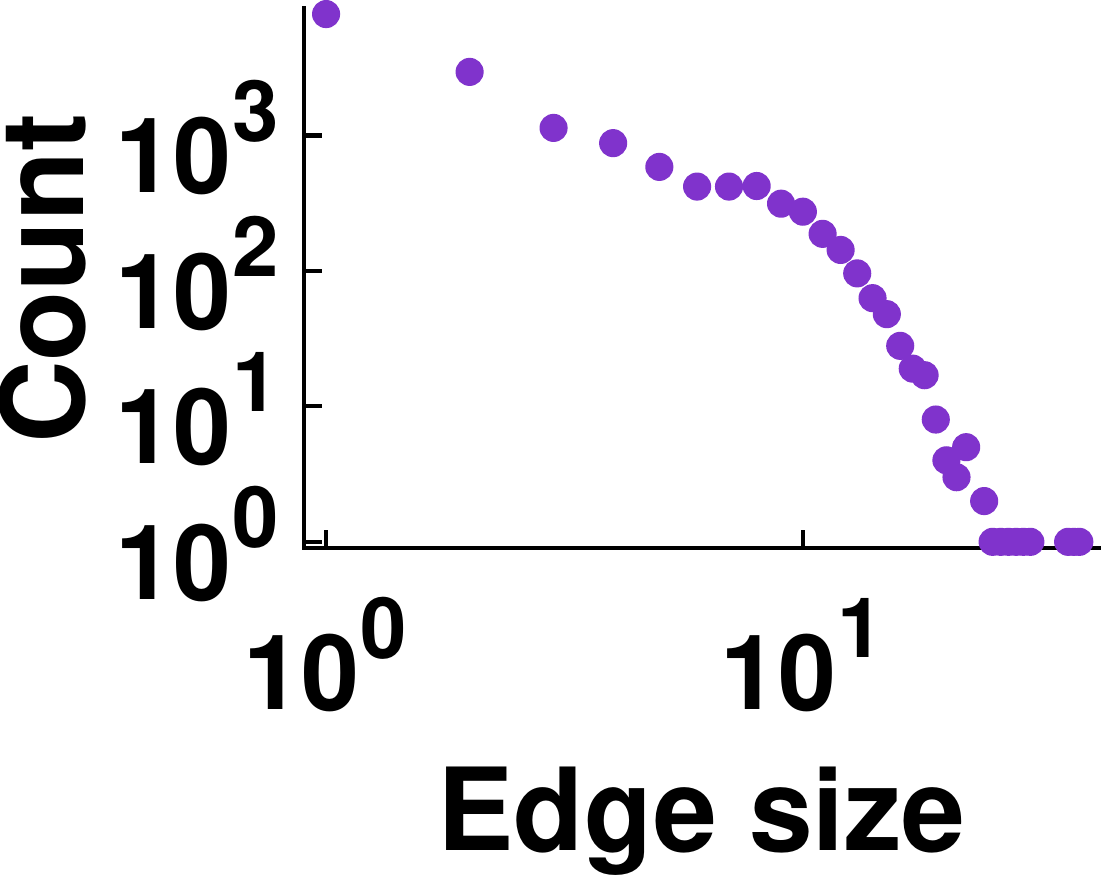} &
			\includegraphics[height=0.762in]{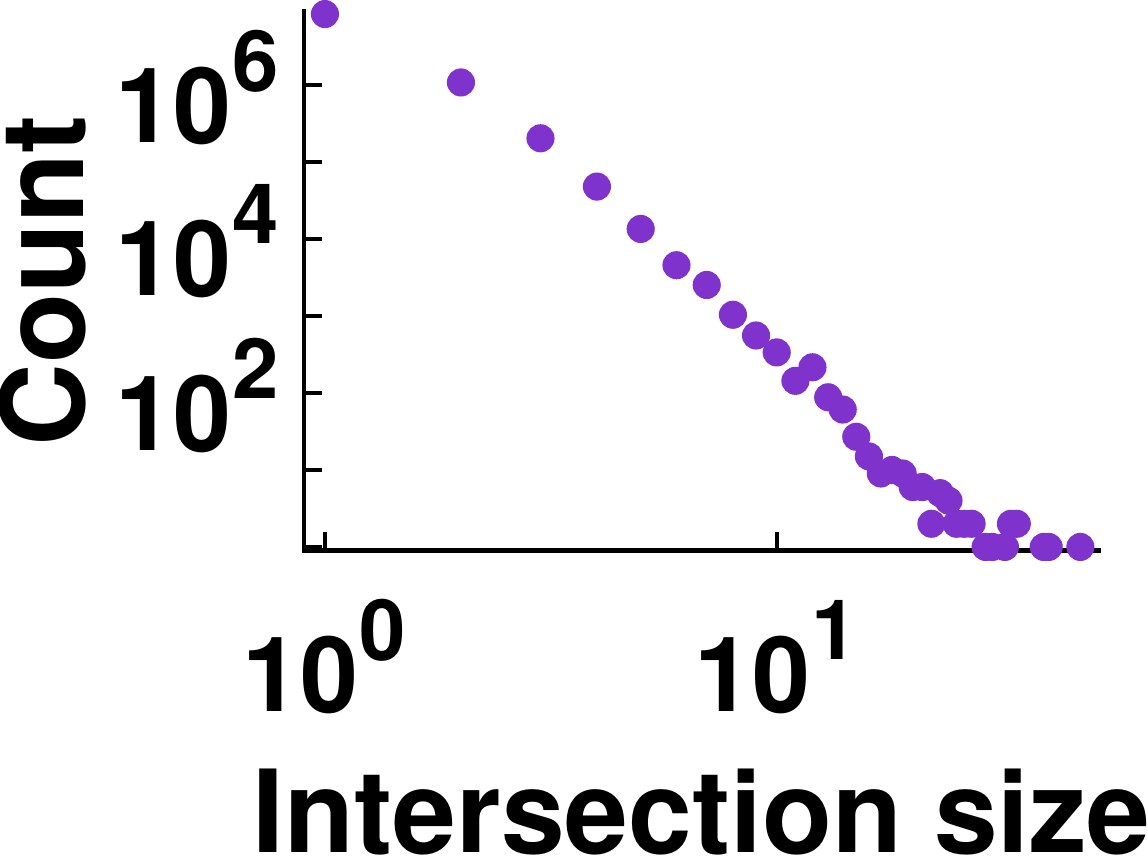} &
			\includegraphics[height=0.762in]{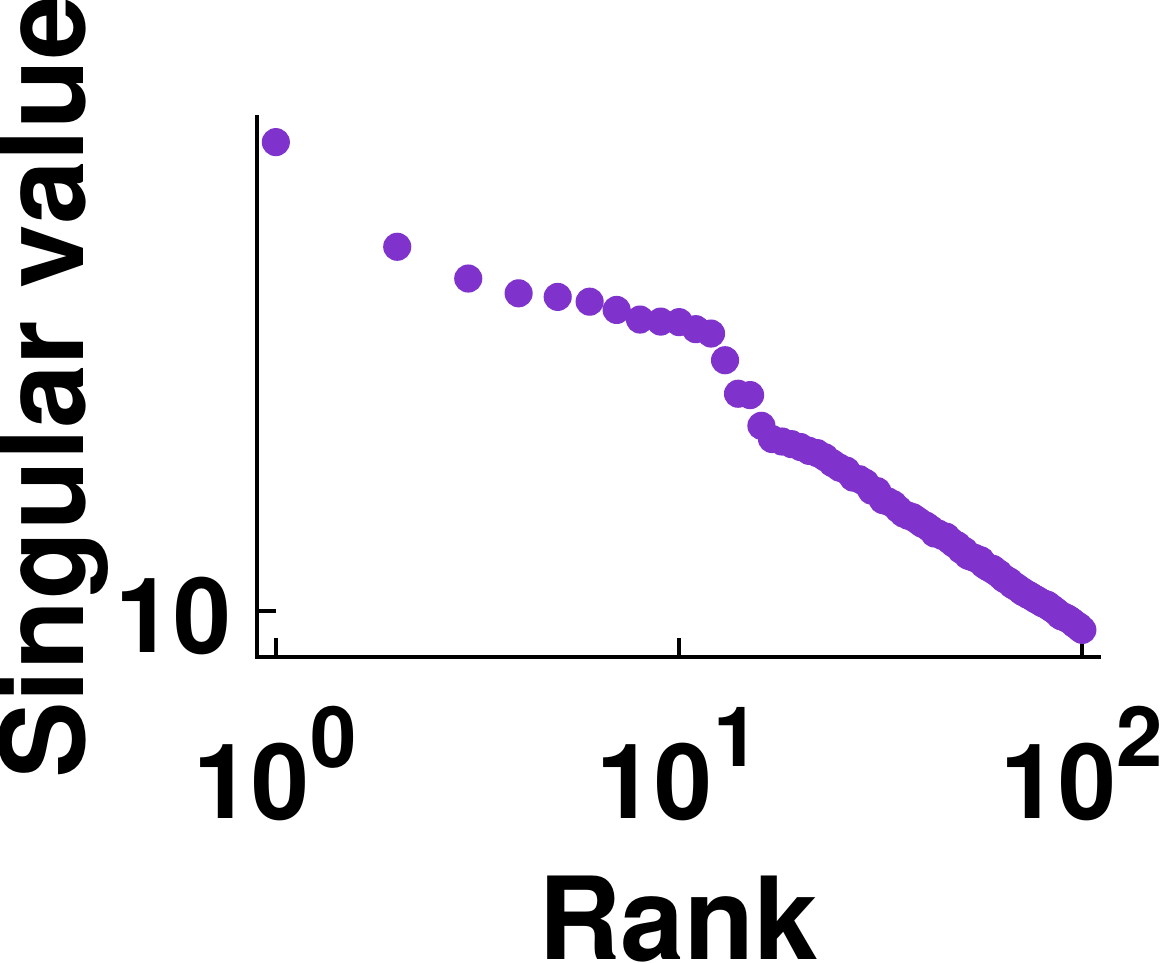} &
			\includegraphics[height=0.762in]{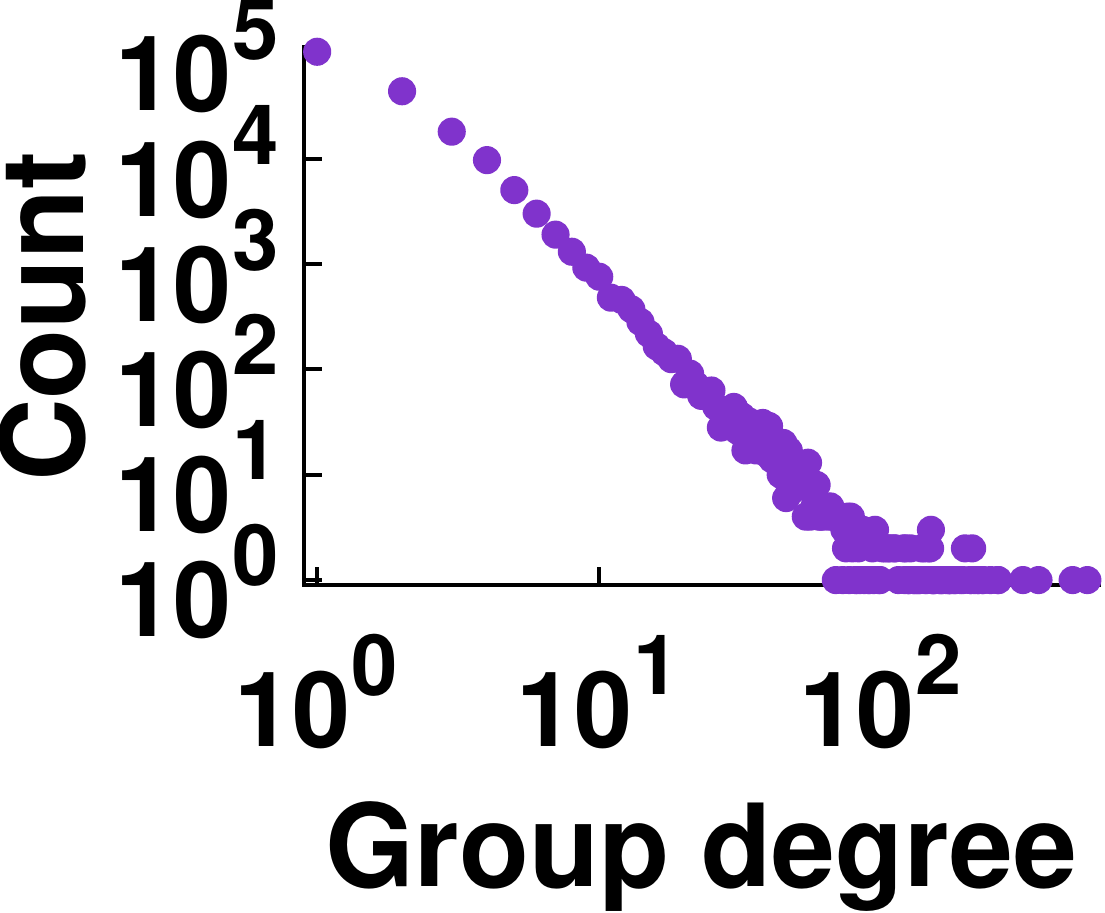} &
			\includegraphics[height=0.762in]{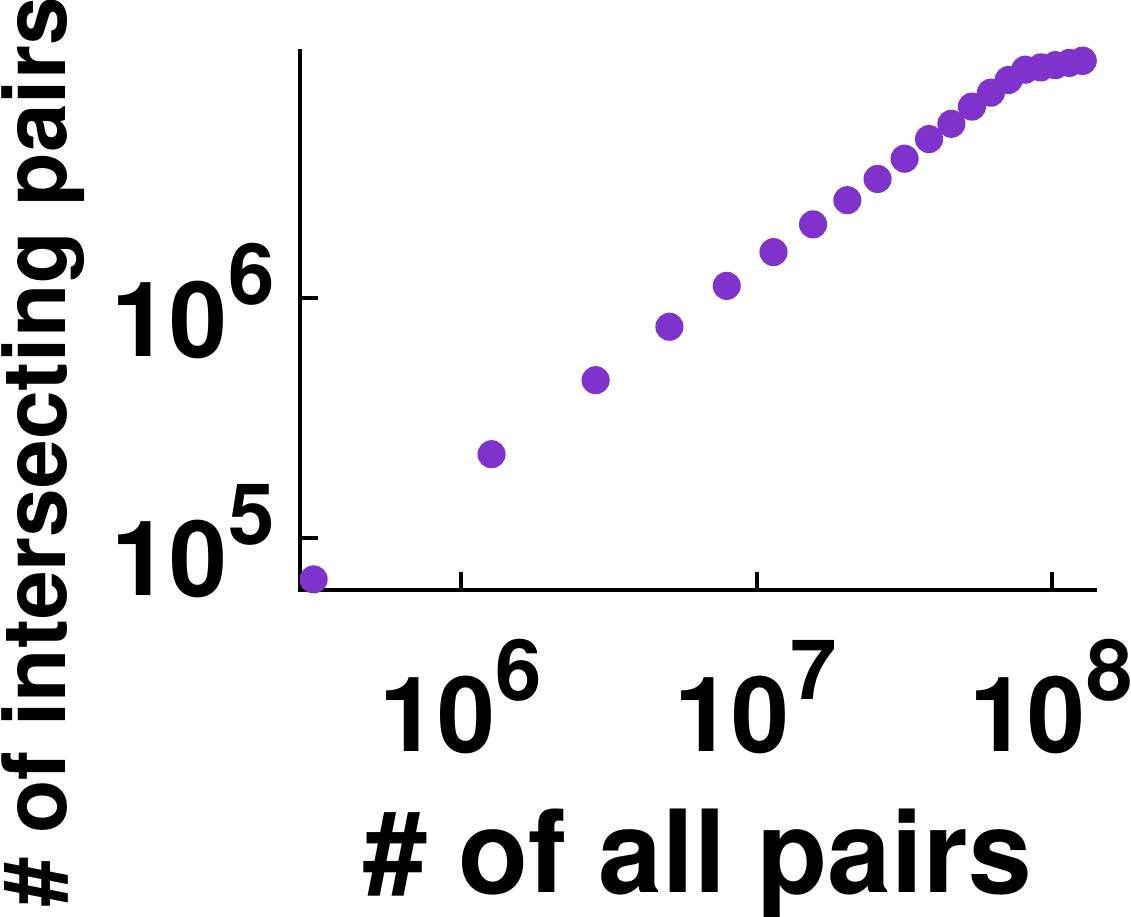} &
			\includegraphics[height=0.762in]{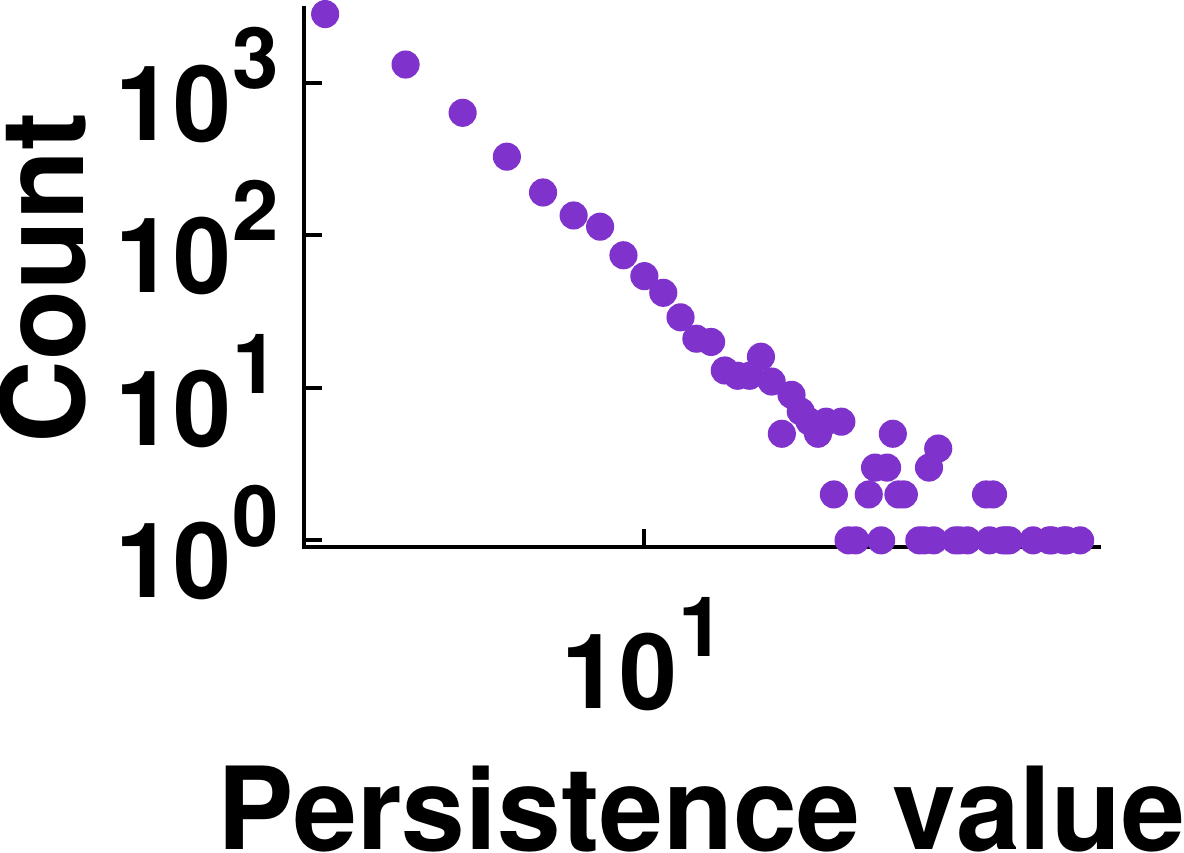}
			\\
			\rotatebox[origin=l]{90}{(P = 0.75)} &
			
			\includegraphics[height=0.762in]{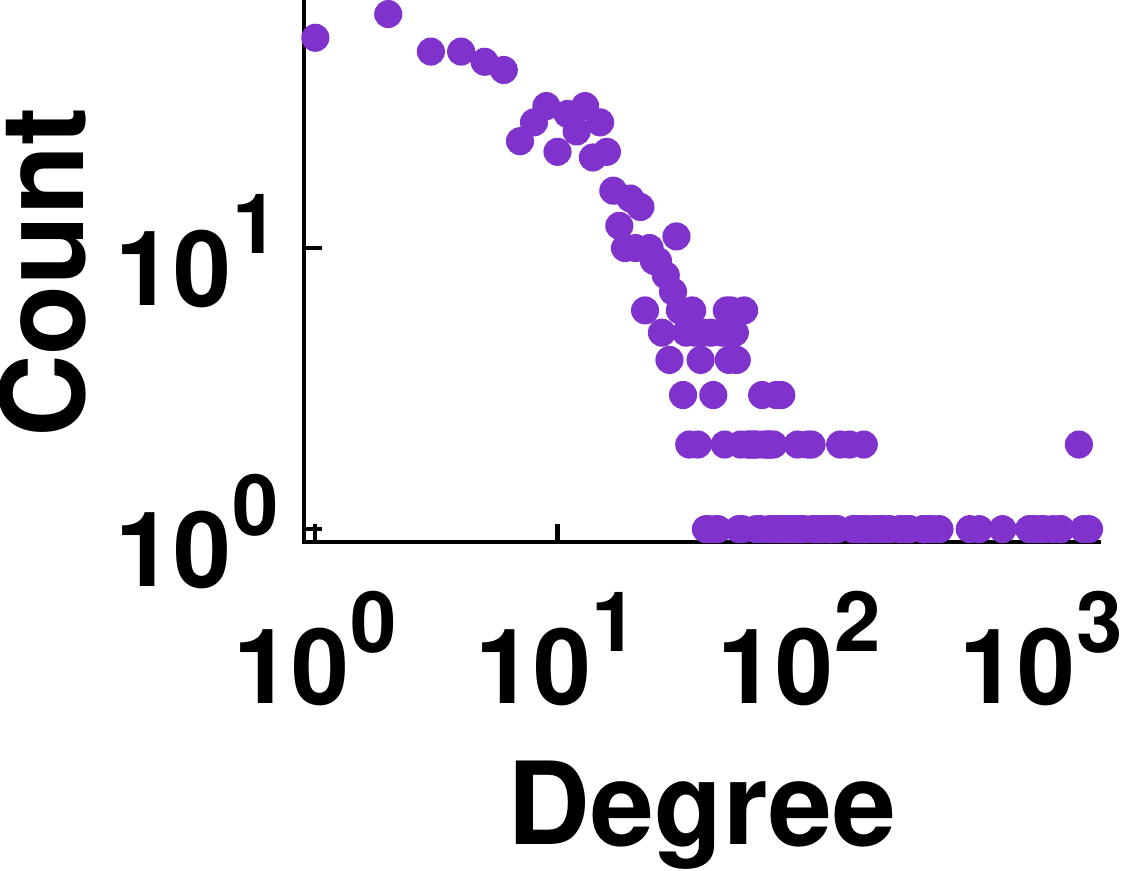} &
			\includegraphics[height=0.762in]{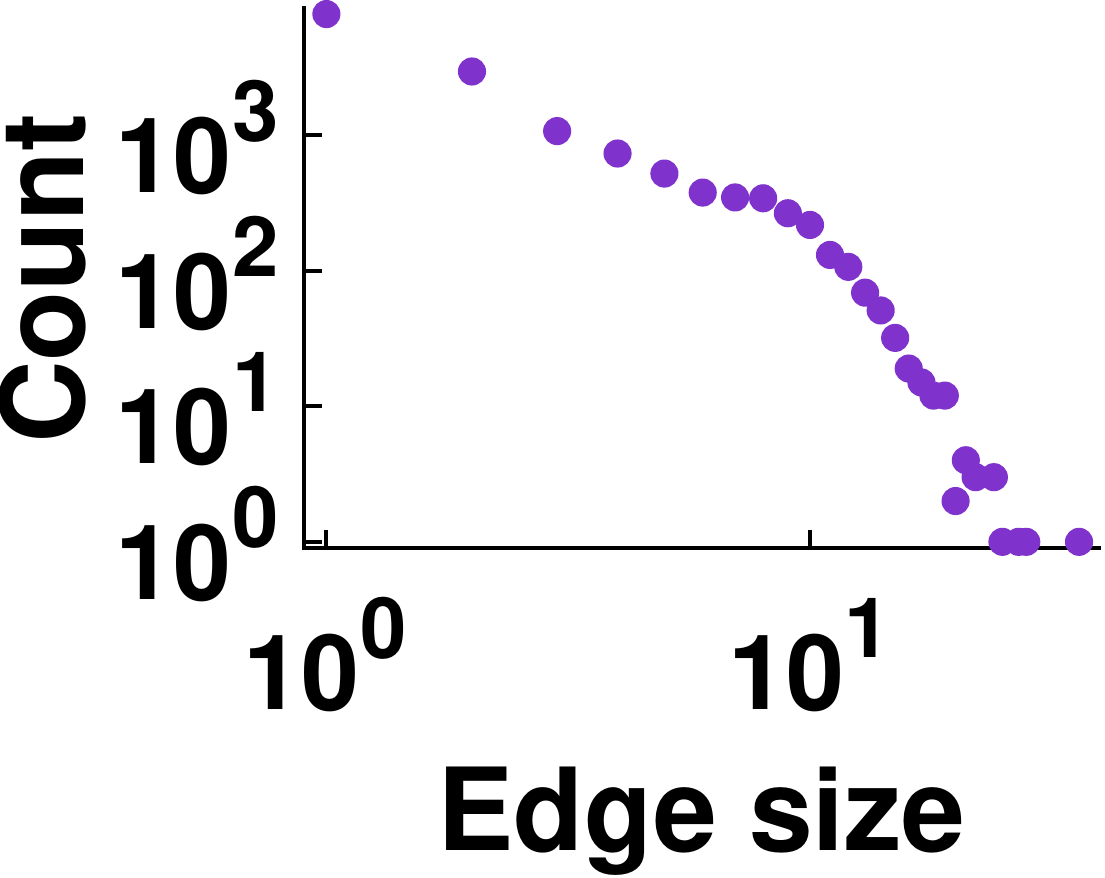} &
			\includegraphics[height=0.762in]{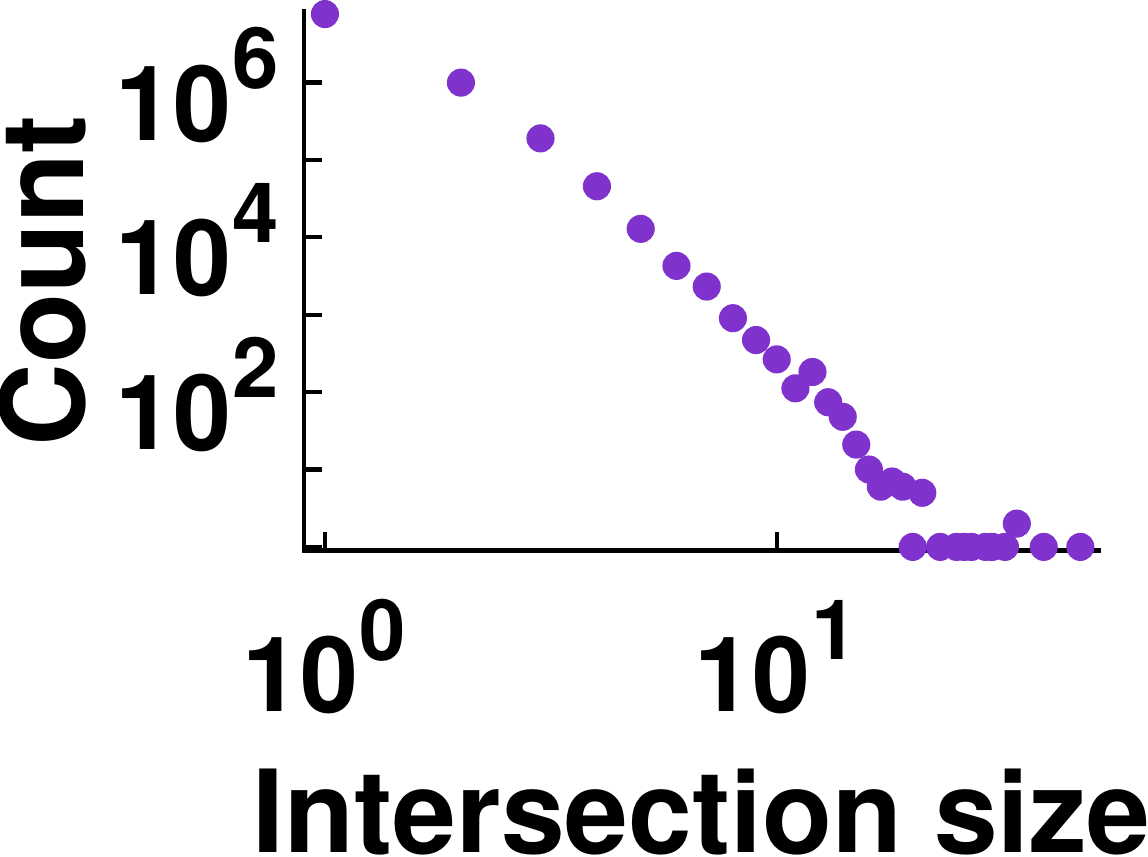} &
			\includegraphics[height=0.762in]{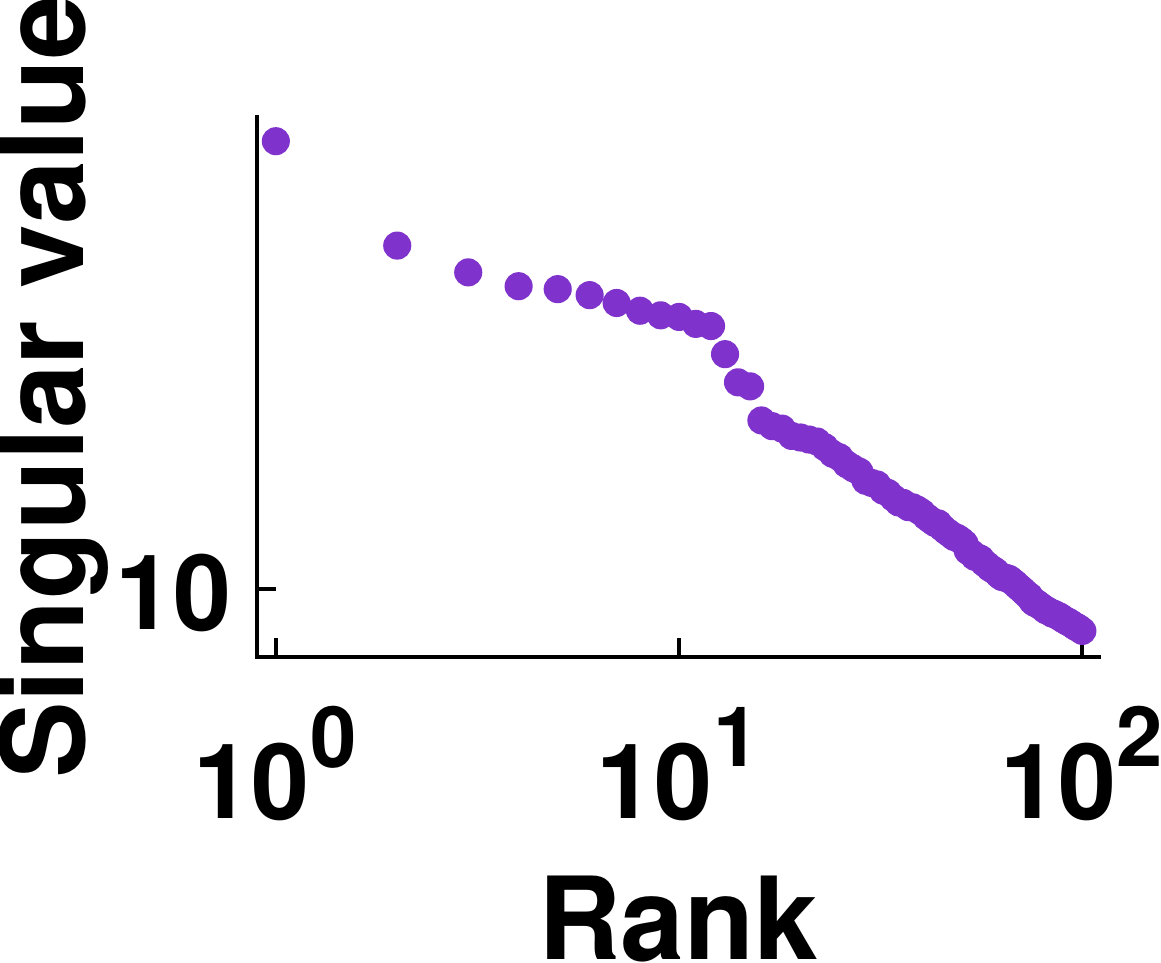} &
			\includegraphics[height=0.762in]{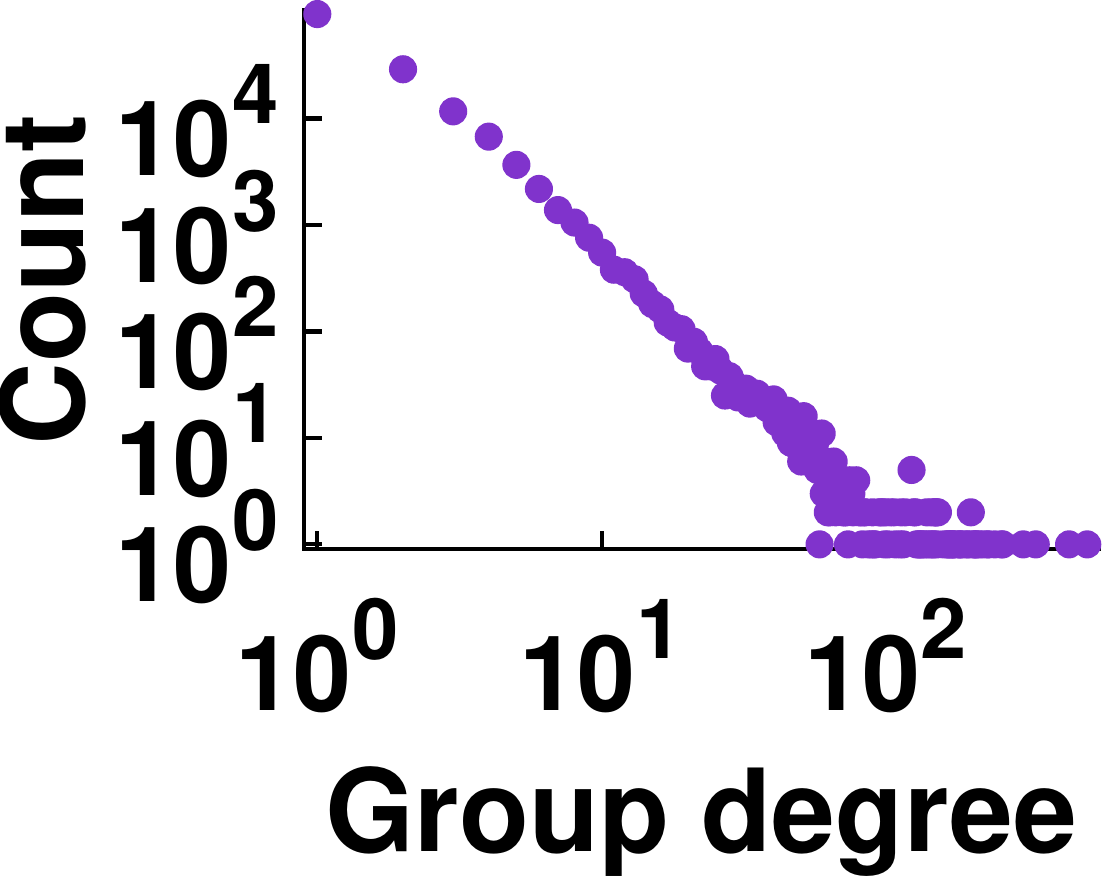} &
			\includegraphics[height=0.762in]{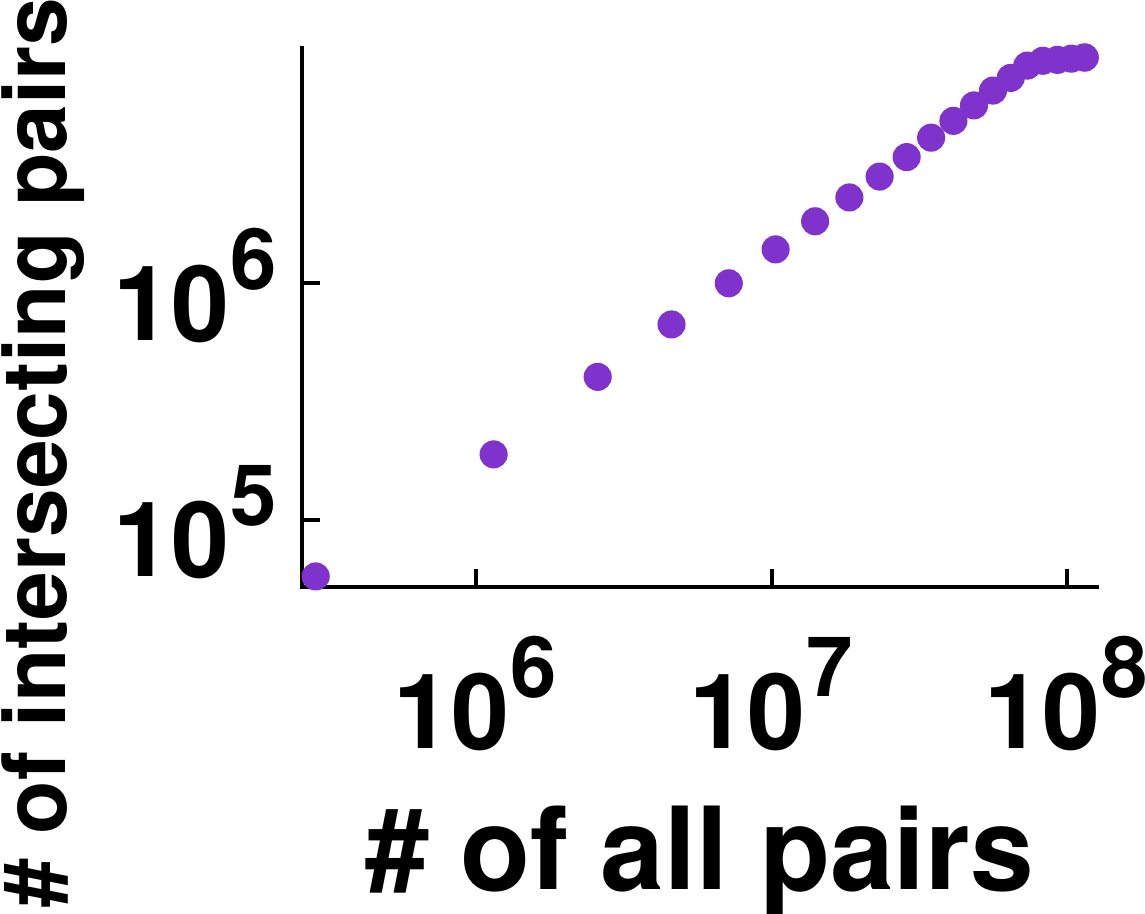} &
			\includegraphics[height=0.762in]{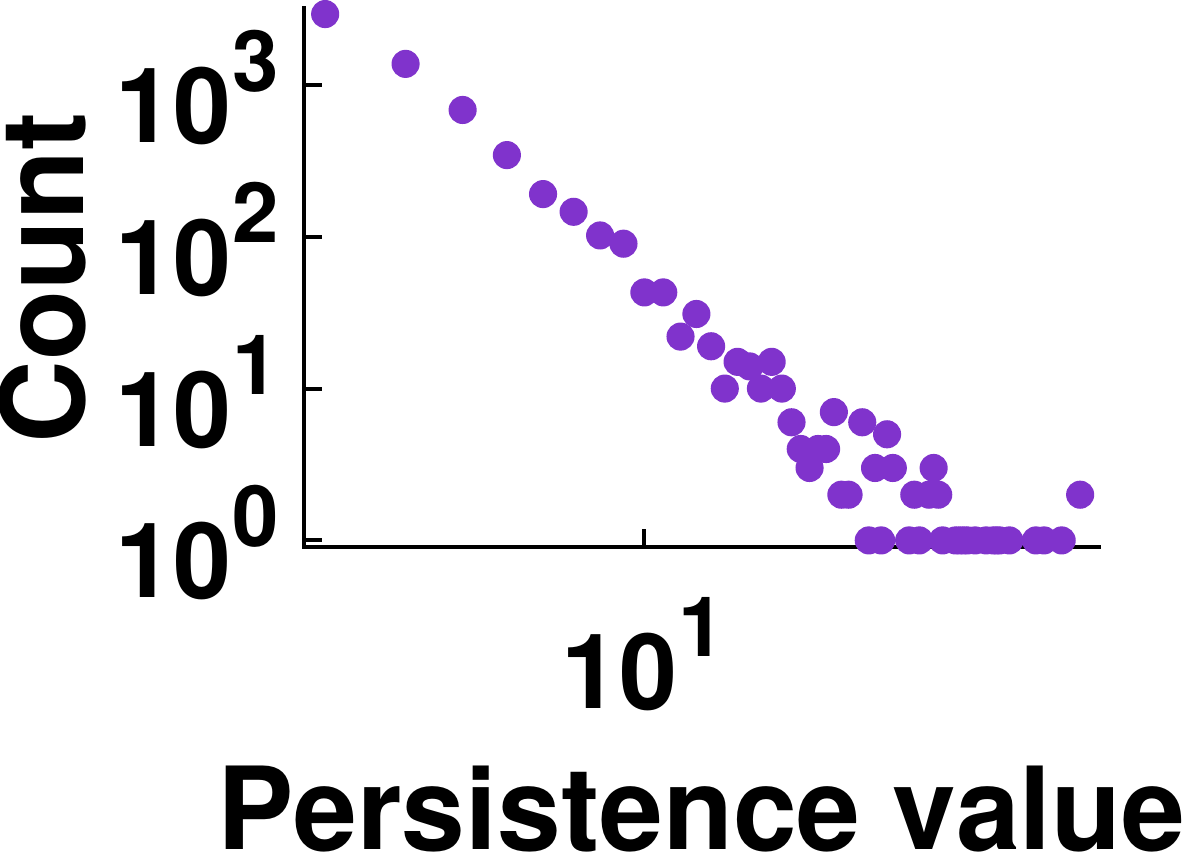}
			\\
			\rotatebox[origin=l]{90}{(P = 0.85)} &
			
			\includegraphics[height=0.762in]{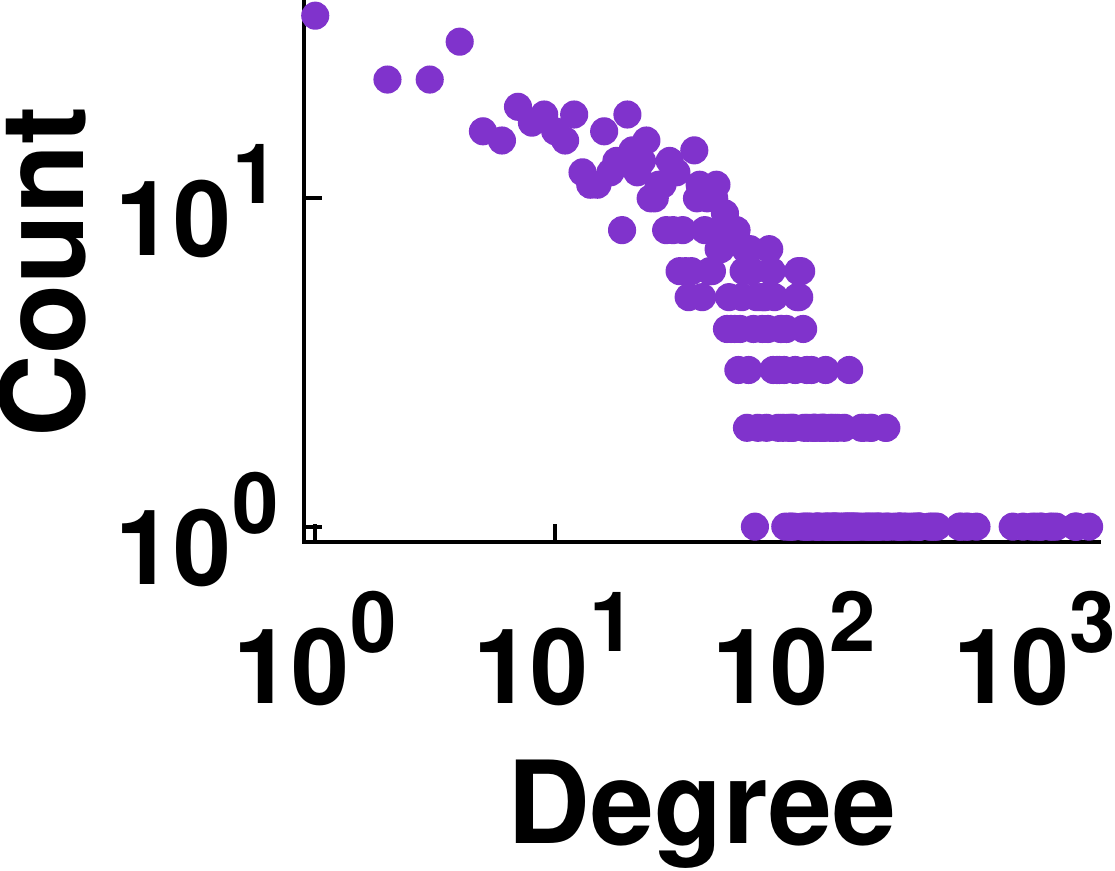} &
			\includegraphics[height=0.762in]{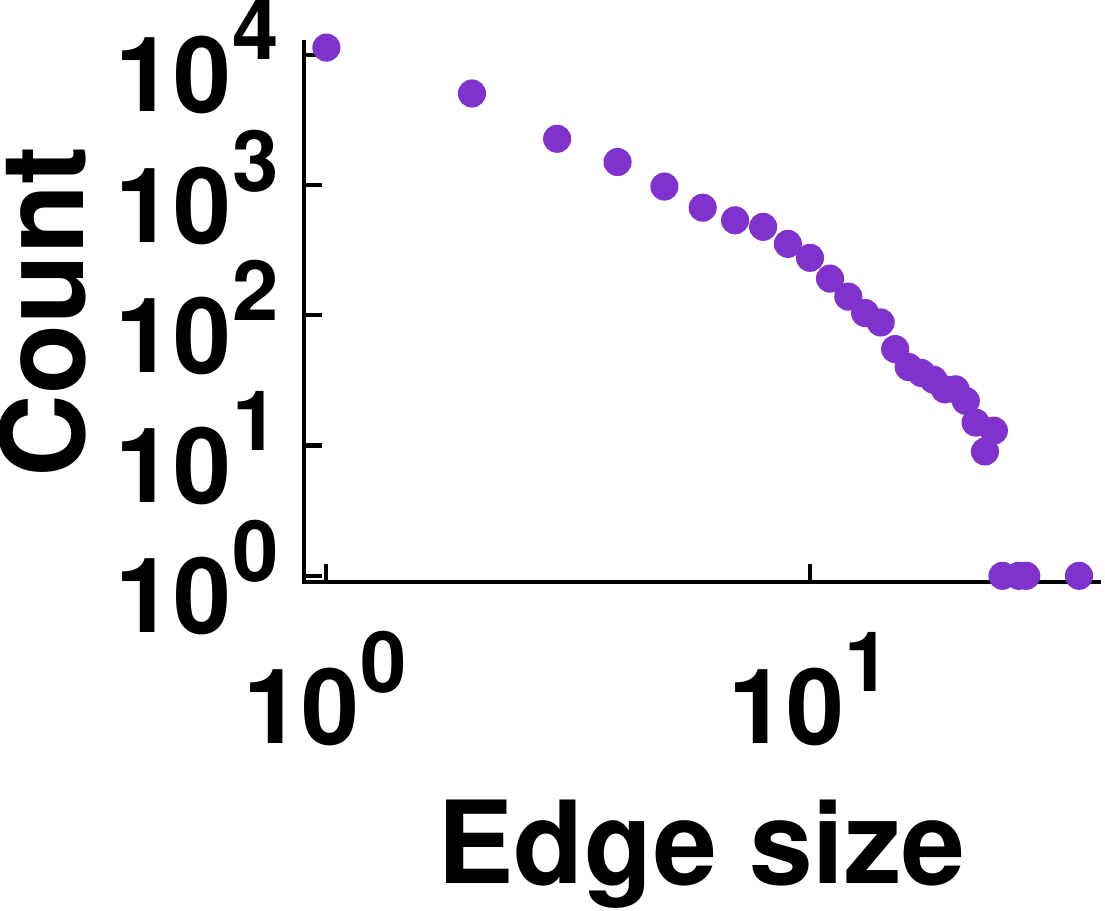} &
			\includegraphics[height=0.762in]{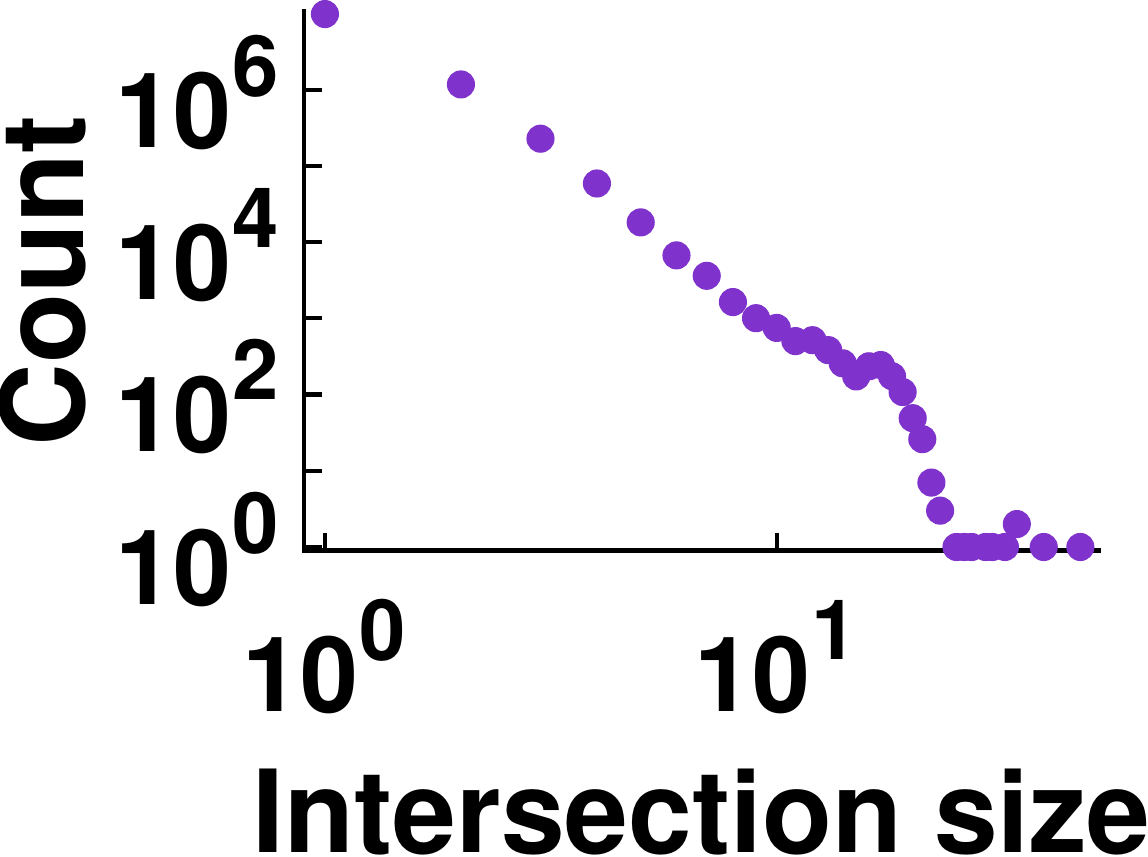} &
			\includegraphics[height=0.762in]{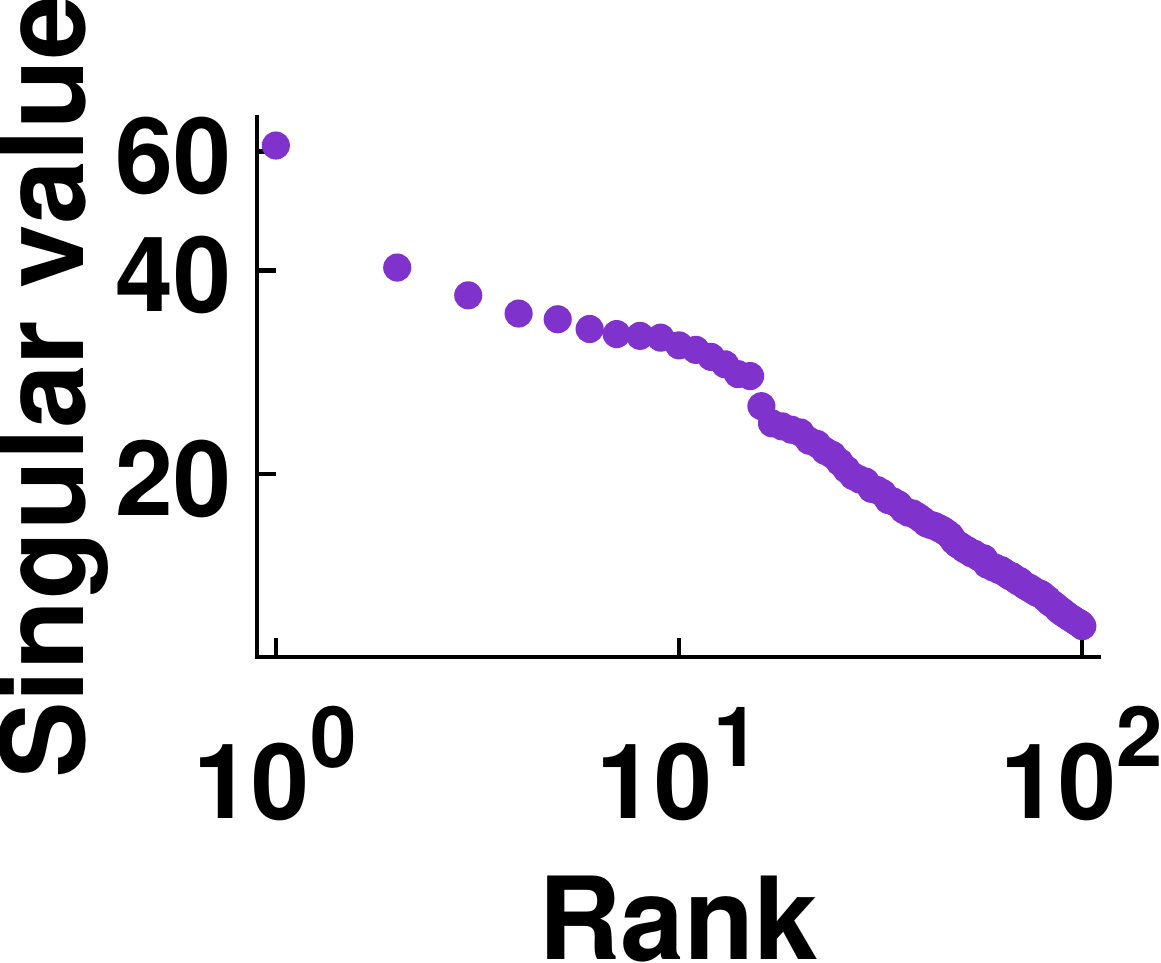} &
			\includegraphics[height=0.762in]{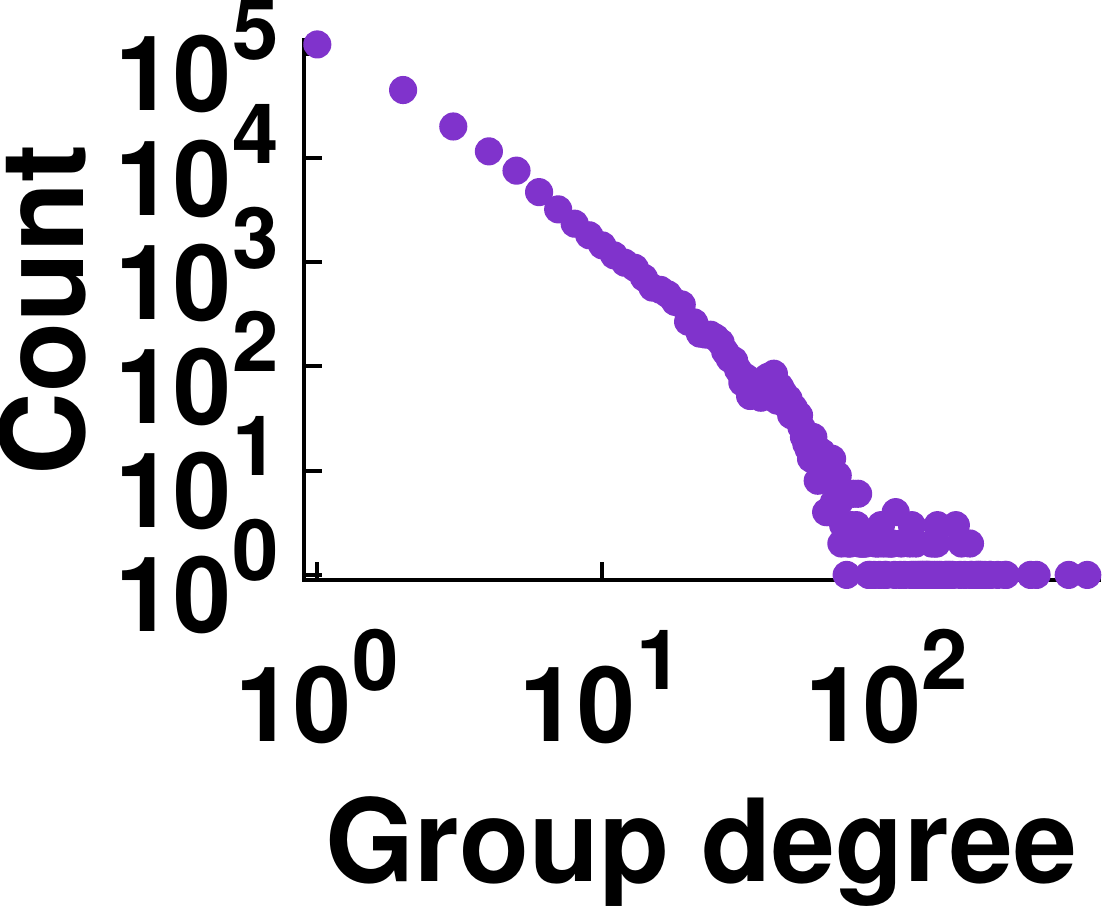} &
			\includegraphics[height=0.762in]{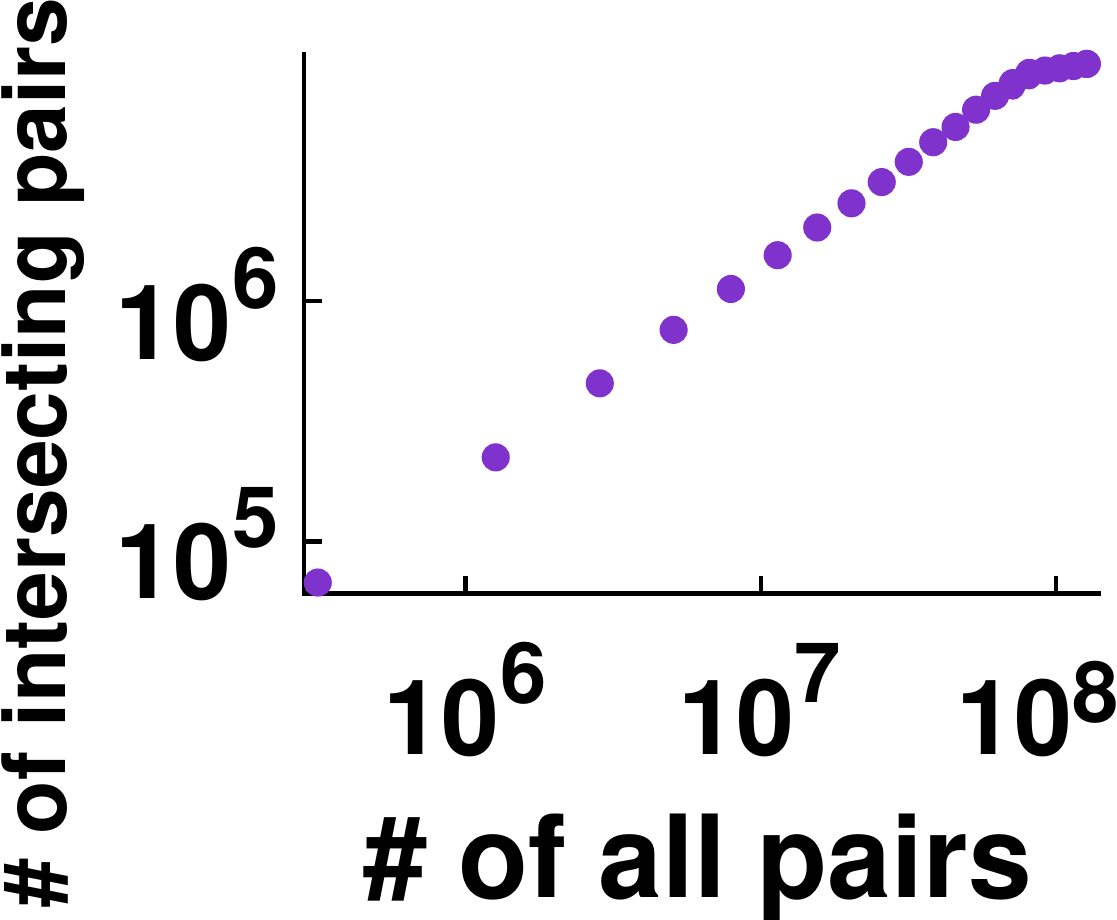} &
			\includegraphics[height=0.762in]{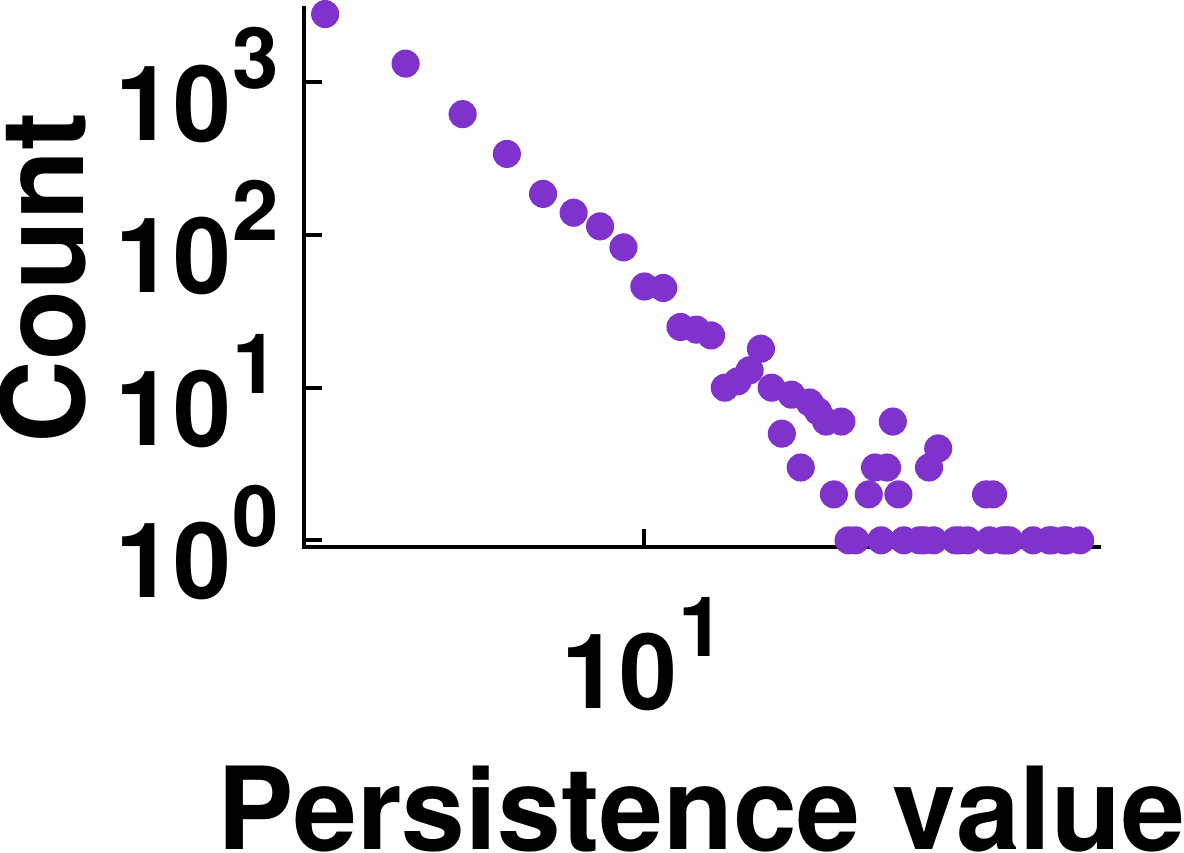}
			\\
			\cmidrule(lr){2-6}
			\cmidrule(lr){7-8}
			\rotatebox[origin=l]{90}{Real Data}
			&
			\includegraphics[height=0.762in]{figures/Dfig/email-1} &
			\includegraphics[height=0.762in]{figures/Dfig/email-2} &  	
			\includegraphics[height=0.762in]{figures/Dfig/email-3} &
			\includegraphics[height=0.762in]{figures/Dfig/email-4} &
			\includegraphics[height=0.762in]{figures/Dfig/email-6} &
			\includegraphics[height=0.762in]{figures/Dfig/email-5} &
			\includegraphics[height=0.762in]{figures/Dfig/email-8}
			\\
			\bottomrule
		\end{tabular}
	}
\end{figure*}

In this section, we conduct a comprehensive experimental evaluation to validate the effectiveness of our HyperLLM framework. We first compare its performance against state-of-the-art baselines in fitting real-world hypergraph structures. We then analyze the sensitivity of its key hyperparameters to demonstrate its robustness.

\begin{figure*}[t!]
	\centering
	\caption{
		\textbf{Hyperparameter sensitivity analysis of HyperLLM.} The plots show performance on four datasets: (a) Email-Eu-core, (b) coauth-Geo, (c) DAWN, and (d) NDC-substances. The z-axis ($\gamma$) measures goodness of fit ($\gamma$, defined in Eq.~\ref{eq:gamma}); higher is better.
	}
	\label{fig:hyper_params_3d_new}
	\begin{subfigure}[b]{0.24\textwidth}
		\centering
		\includegraphics[width=\textwidth]{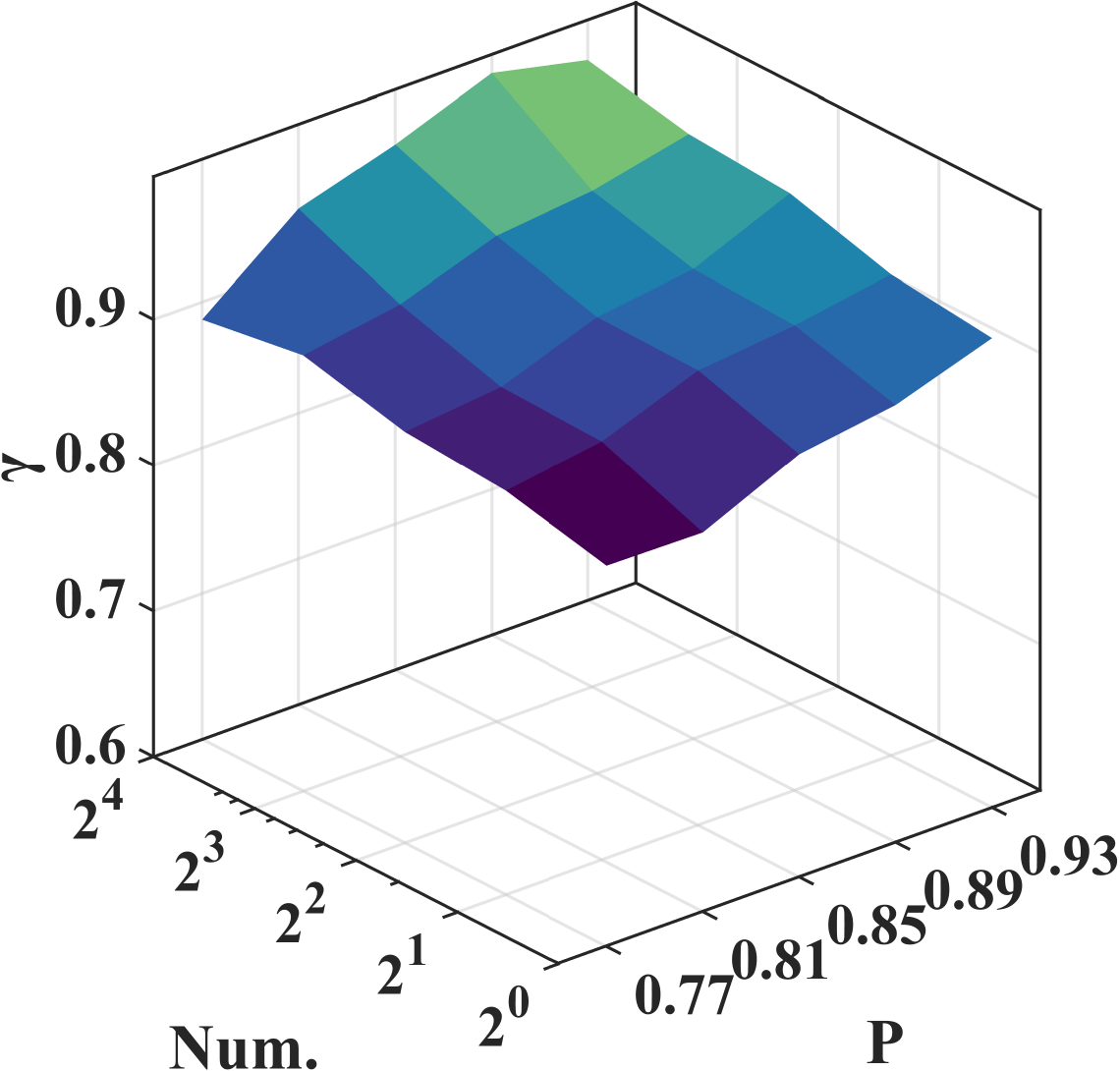}
		\caption{Email-Eu-core}
	\end{subfigure}
	\hfill
	\begin{subfigure}[b]{0.24\textwidth}
		\centering
		\includegraphics[width=\textwidth]{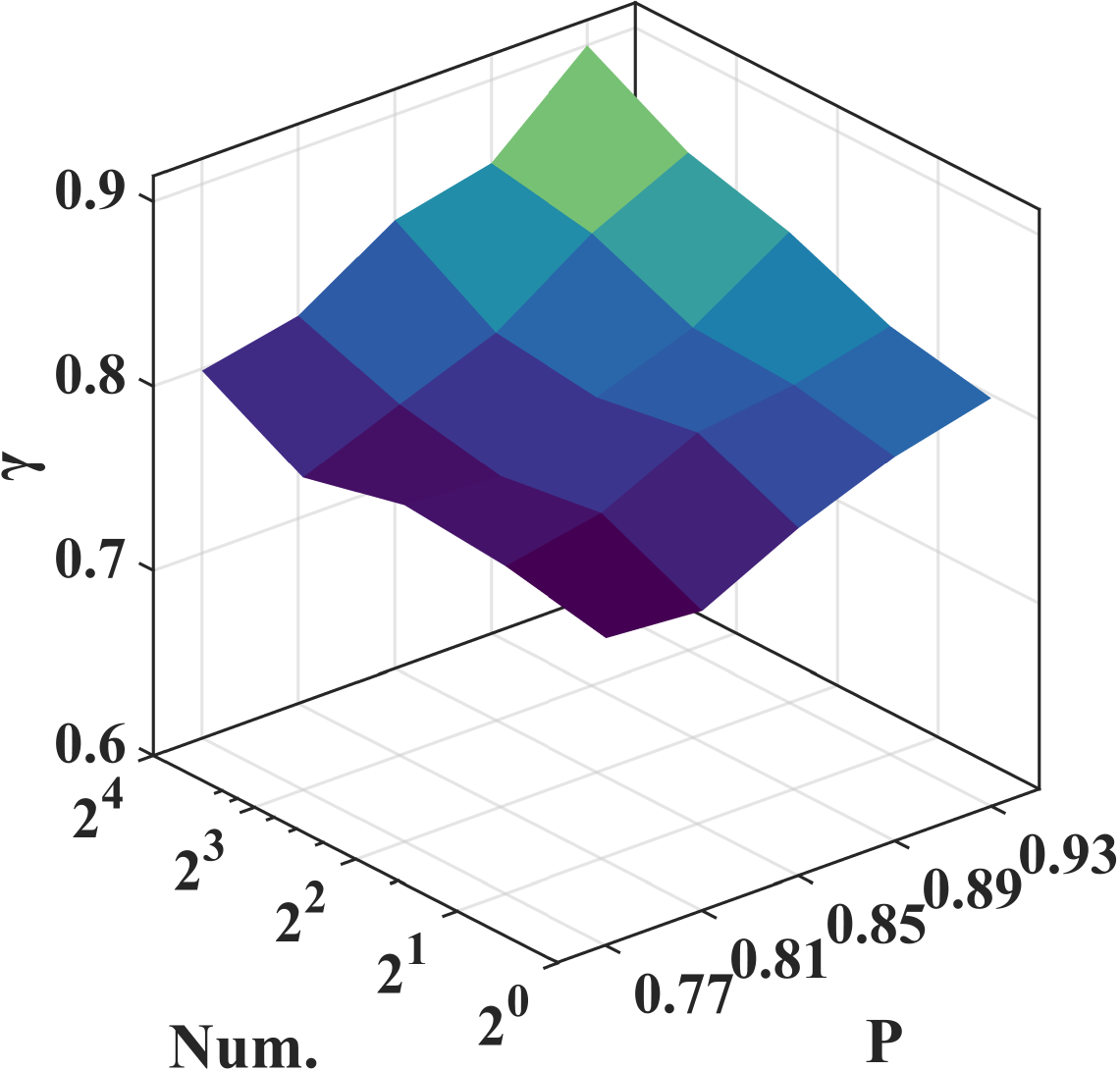}
		\caption{coauth-Geo}
	\end{subfigure}
	\hfill
	\begin{subfigure}[b]{0.24\textwidth}
		\centering
		\includegraphics[width=\textwidth]{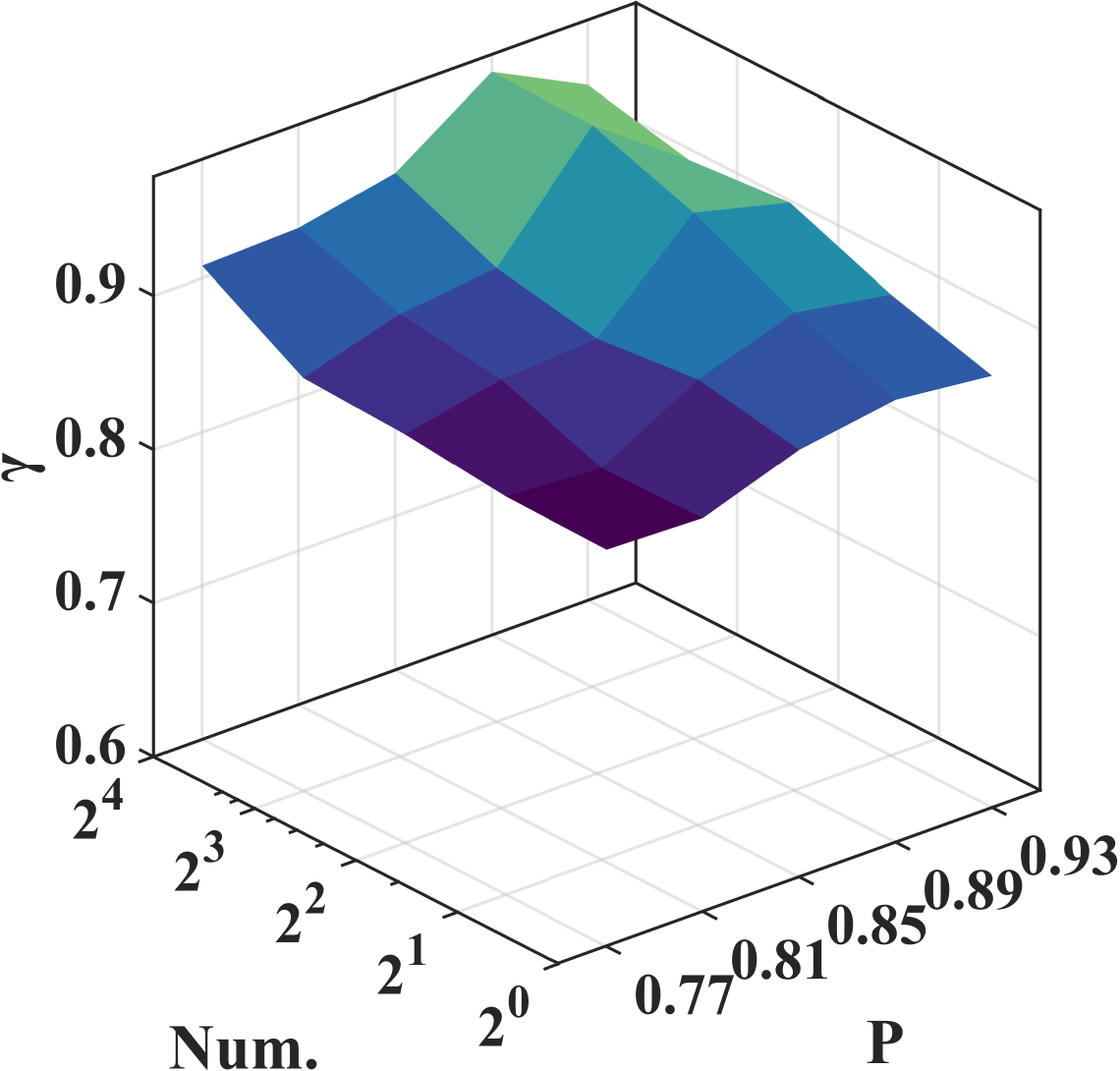}
		\caption{DAWN}
	\end{subfigure}
	\hfill
	\begin{subfigure}[b]{0.24\textwidth}
		\centering
		\includegraphics[width=\textwidth]{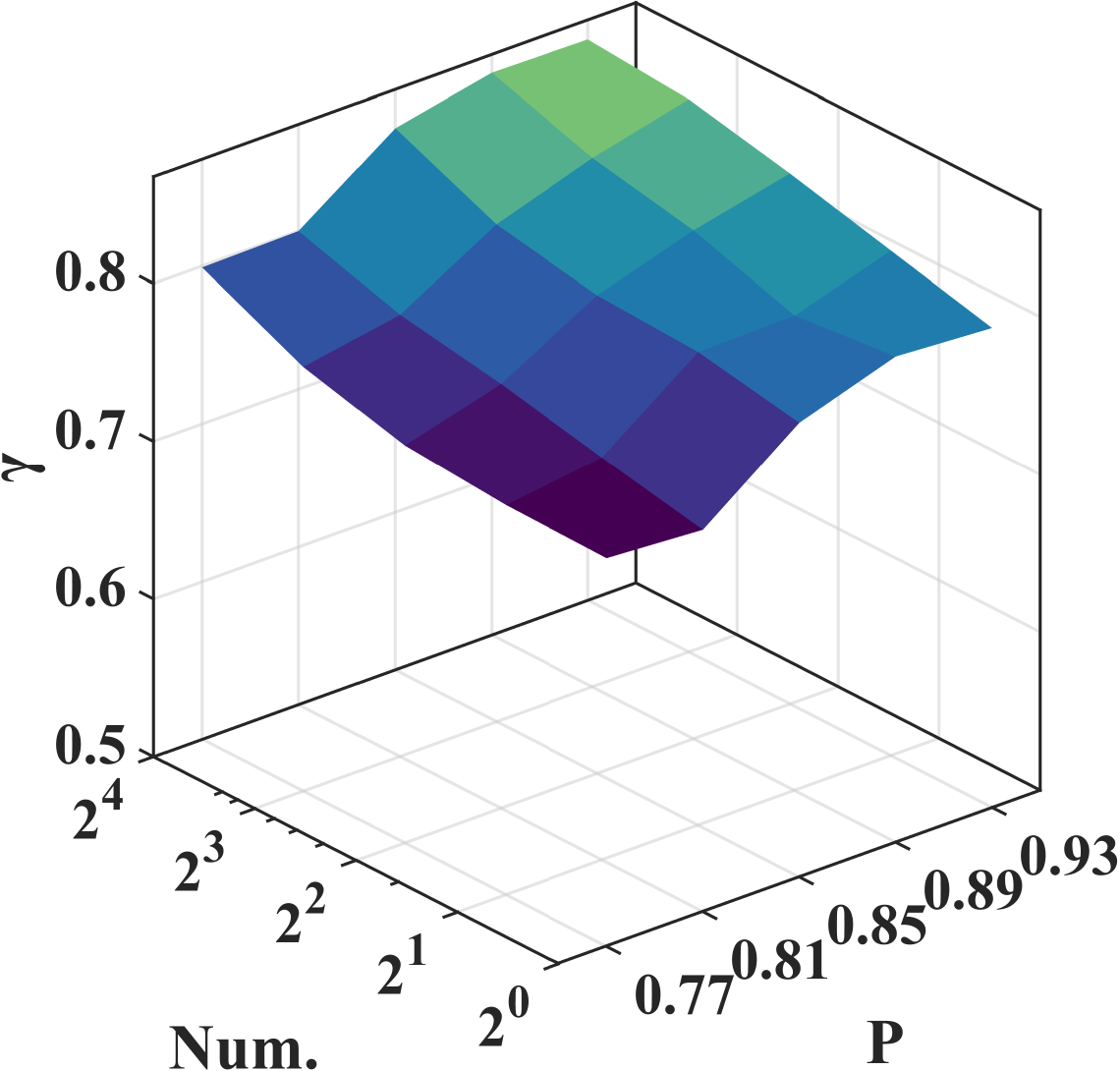}
		\caption{NDC-substances}
	\end{subfigure}
\end{figure*} 

\subsection{Comparative Analysis with Baselines}

To assess HyperLLM's ability to capture complex structures, we evaluate it against a suite of strong baseline models on eight diverse, real-world datasets. The results, summarized in Table~\ref{tab:llm_fitting}, highlight the superiority of our approach. HyperLLM achieves the best average ranking across the eight structural and dynamic patterns identified in Section~4, outperforming all baselines. 

Notably, this top-tier performance is achieved with minimal input requirements. As shown in the "Input Information" columns of Table~\ref{tab:llm_fitting}, most conventional generators rely on explicit structural properties of the target graph, such as the number of nodes (NN), the full degree distribution (DD) or size distribution (SD). In contrast, HyperLLM requires no such prior structural information, leveraging the rich semantic attributes of the nodes to drive its generation process. The size distribution (SD) is marked as an optional input (\textcolor{red}{?}); we provide it only to guide the model's construction phase for efficiency on a large scale, but it is not fundamental to the model's generative logic. This demonstrates a key advantage of our framework: its ability to produce structurally faithful hypergraphs by modeling the underlying agent interactions, rather than merely replicating statistical patterns.

Figure~\ref{fig:model_params} provides a qualitative validation of HyperLLM's fitting capabilities. The figure compares seven key patterns from a real-world dataset (bottom row) against those generated by HyperLLM using four different preferential attachment probabilities. A clear visual correspondence emerges as the attachment probability increases, with the patterns generated using a probability of 0.85 closely mirroring the ground truth. This alignment not only confirms the model's effectiveness but also empirically supports the theoretical preferential attachment mechanism discussed in Section~4.

\subsection{Hyperparameter Sensitivity}

We further analyze the sensitivity of HyperLLM's two primary hyperparameters: the preferential attachment probability during the iterative construction phase and the number of entities suggested by the Optimizer agent during the multi-agent evolution phase. To quantify the goodness of fit, we define a metric, $\gamma$, as:
\begin{equation}
\label{eq:gamma}
    \gamma = 1 - \frac{|\text{slope}_{\text{real}} - \text{slope}_{\text{exp}}|}{|\text{slope}_{\text{real}}|}
\end{equation}
where $\text{slope}_{\text{real}}$ and $\text{slope}_{\text{exp}}$ are the power-law exponents of the degree distributions for the real and generated hypergraphs, respectively. A value of $\gamma$ closer to 1 indicates a more accurate fit.

Figure~\ref{fig:hyper_params_3d_new} visualizes the performance landscape across four different datasets. The results show that HyperLLM is robust, with performance varying smoothly across a wide range of parameter settings rather than being confined to a narrow optimal region. Generally, the best results are achieved with a high attachment probability and a moderate number of entities suggested by the Optimizer. This suggests a synergistic relationship between the two core mechanisms: strong preferential attachment establishes a rich-get-richer backbone, while the multi-agent system provides the nuanced, localized refinements needed to capture higher-order structural details.
 

%% file: Body/7.tex
\section{Conclusions}

In this paper, we introduced HyperLLM, a novel framework pioneering the use of Large Language Models for generating realistic, high-order hypergraphs. By modeling the principles of human interaction, our approach moves beyond statistical replication to a semantic, behavior-driven synthesis. Our work shows that LLMs, guided by a multi-agent system and grounded in network science theory, create synthetic hypergraphs with remarkable fidelity to their real-world counterparts.

Our primary contributions and findings can be summarized as follows:
\begin{itemize}[leftmargin=*,noitemsep,topsep=0pt]
    \item We proposed a new paradigm for hypergraph generation that leverages the semantic reasoning of LLMs. Our experimental results show that HyperLLM significantly outperforms established baselines, achieving superior structural accuracy while requiring minimal prior information about the target network's statistics.
    \item We designed a flexible, dual-phase generation framework. It combines a rapid iterative algorithm for initial construction with a sophisticated multi-agent system (MAS) for evolutionary refinement, offering a versatile toolkit for different generation needs.
    \item We provided a new theoretical explanation for the prevalence of heavy-tailed patterns in hypergraphs by extending a microscopic dynamic model, thereby enhancing the interpretability of the generation process.
\end{itemize}

This research opens several avenues for future work. A key direction is to expand the scope of evolution from hyperedges to a fully dynamic system where nodes can also join or leave. Another promising path is the generation of more complex structures, such as heterogeneous or signed hypergraphs, to model a wider range of real-world systems. Furthermore, while our work lays the groundwork for agent-based generation, future studies could delve deeper into the emergent communication and collaboration strategies among autonomous LLM agents \cite{hong2023metagpt, qian2023communicative}, potentially using techniques like reinforcement learning to train even more sophisticated agent behaviors \cite{hao2023improving}, thus offering new insights into computational social science.

Finally, exploring hybrid models that couple the semantic intelligence of LLMs with algorithmic efficiency could unlock massive-scale network generation, pushing the frontiers of AI-assisted network science. 